\documentclass[showpacs,preprint,preprintnumbers,amsmath,amssymb,author-numerical,aip,cha]{revtex4-1}

\usepackage{graphicx}
\usepackage{dcolumn}
\usepackage{bm}
\usepackage{multirow}
\usepackage{url}
\usepackage{color}

\usepackage{ifpdf}

\begin{document}

\preprint{}

\title{Detecting chaos in particle accelerators through the frequency map analysis method}

\author{Yannis Papaphilippou}%
 \email{Ioannis.Papaphilippou@cern.ch}
\affiliation{European Organisation of Nuclear Research - CERN, Geneva, Switzerland}%

\date{\today}

\begin{abstract}
 The motion of beams in particle accelerators is dominated by a plethora of non-linear effects which can enhance chaotic motion and limit their performance. The application of advanced non-linear dynamics methods for detecting and correcting these effects and thereby increasing the region of beam stability plays an essential role during the accelerator design phase but also their operation. After describing the nature of non-linear effects and their impact on performance parameters of different particle accelerator categories, the theory of non-linear particle motion is outlined. The recent developments on the methods employed for the analysis of chaotic beam motion are detailed. In particular, the ability of the frequency map analysis method to detect chaotic motion and guide the correction of non-linear effects is demonstrated in particle tracking simulations but also experimental data.
 \end{abstract}

\pacs{Valid PACS appear here}
\maketitle

\begin{quotation}

Due to their collective
nature and the external electro-magnetic fields employed for guiding and focusing
them, the dynamics of beams in particle accelerators is intrinsically  non-linear. 
The performance of a wide spectrum of accelerators, ranging from ultra-high energy hadron colliders
with several kilometres of circumference to X-ray storage rings and high-power synchrotrons, can be severely
limited by the resulting chaotic motion, impacting the machine availability but also construction
and operating cost. Methods based on high order perturbation theory
have found ample room for application in beam dynamics. 
A breakthrough in the field was the introduction of the Frequency
Map Analysis method~\cite{ref:NAFF1,ref:NAFF2,ref:NAFF3} which enables the analysis of multi-dimensional Hamiltonian systems
and the clear distinction between ordered and chaotic trajectories, which are either computed by direct numerical integration
or experimentally measured in operating accelerators. This paper reviews  the modern  trends in  detecting chaos
in particle accelerators giving emphasis to the application of frequency map analysis in a variety of beam dynamics
problems.

\end{quotation}

\section{\label{sec:level1}Introduction\protect\\ }

The last few years were remarkable for CERN and the scientific community of particle physics
with the discovery of the Higgs boson by the two experiments ATLAS~\cite{bib:HiggsATLAS} and 
CMS~\cite{bib:HiggsCMS}, which culminated
with the attribution of the 2013 Nobel prize in physics to Fran{\,c}ois Englert  and Peter Higgs~\cite{bib:Nobel}. 
These outstanding experimental
results were also due to the excellent performance of the Large Hadron Collider (LHC), a superb scientific
instrument for particle physics research. 

Modern accelerator rings as the LHC are pushing their design and operating parameters into regimes 
of extreme beam currents and small phase space beam volumes (the so called emittance),
the ratio of which roughly represents the beam brightness, a measure of accelerator performance.
In these high-brightness regimes, non-linear effects become predominant and the motion of particles  at large amplitudes 
becomes chaotic. This can lead to physical increase of the beam size and to particle losses, which do not only affect  
performance but can also be harmful for  machine equipments and reduce the accelerator availability.
The application of advanced non-linear dynamics methods
during the design but also operation of accelerators is thus important for studying and ensuring the 
long-term stability of particle motion. These studies have to guide the design of accelerator components
providing tolerances for the quality of magnetic fields, their alignment, the stability of the power
supplies, the use of corrector magnets adequately grouped for the alleviation of these effects, and all
these, at the smallest possible cost.

In this paper, we review the basics of non-linear beam dynamics in  particle accelerator rings, 
the effects that are responsible for the onset of chaos, the methods used for detecting this chaotic motion 
and how these methods guide accelerator design for increasing the beam stability.  
Particular emphasis is given to the Frequency Map Analysis (FMA)~\cite{ref:NAFF1,ref:NAFF2,ref:NAFF3}, which has been proven
a very robust numerical method for exploring and understanding the global dynamics of any non-linear Hamiltonian system.

The paper is organised as follows:
in section~\ref{sec:basics}, the fundamentals of non-linear particle beam  motion are  briefly recalled, including
the accelerator Hamiltonian, one-turn maps and normal forms. The next section is devoted to the review
of different chaos detection techniques in beam dynamics and in particular the FMA method. The application
of this method in various accelerator rings and non-linear beam dynamics problems is presented in section~\ref{accelFMA}.
Experimental non-linear dynamics techniques are elaborated in section~\ref{expNBD} before closing with the summary.

\section{\label{sec:basics}Non-linear beam dynamics basics\protect\\ }

%
%
%
%
%
%
%
%

%

\subsection{The accelerator Hamiltonian}

In any circular accelerator, particles are moving inside vacuum pipes under 
the influence of self-generated but also external three-dimensional electromagnetic fields.
Electric fields produced by radio-frequency (RF) cavities are usually used 
to accelerate particles and affect mostly longitudinal 
(``synchrotron'') motion, whereas magnetic fields are used for their guidance and dominate
transverse (``betatron") motion. Self-fields are represented by Coulomb forces within the particles 
of the same beam or through image charges in the vacuum chamber or between two closely interacting beams
and dominate collective beam motion.

Neglecting self fields and radiation, the system can be described by a Òsingle-particleÓ Hamiltonian~\cite{Jackson}
\begin{equation}
H({\bf x},{\bf p},t) = c \sqrt{\left({\bf p} - 
\frac{e}{c} {\bf A}({\bf x},t) \right)^2 +
m^2c^2} + e \Phi({\bf x},t)
\;\;,
\label{eq:SPRHamp}
\end{equation}
where ${\bf x}=(x,y,z)$ are the usual Cartesian positions, ${\bf p}=(p_x,p_y,p_z)$  their conjugate momenta,
${\bf A}=(A_x,A_y,A_z)$  the magnetic vector potential, 
$\Phi$ the electric scalar potential,  $m$ the particleÕs rest mass, $c$ the speed of light and $e$ the electric charge.  
Using Hamilton's equations $({\bf\dot{x}},{\bf\dot{p}}) = [({\bf x},{\bf
p}),H]$, the usual Lorenz equations
for charged particle motion in electromagnetic fields can be obtained.

The Hamiltonian represents a time-dependent three
degrees of freedom system. 
It is useful (especially for rings) to transform the Cartesian coordinate system to the Frenet-Serret system moving to a closed curve, with path length $s$ and make a further canonical transformation that uses this path length as the independent variable instead of the time $t$. 

The longitudinal motion can be approximated by a time-dependent (periodic) pendulum and is
generally slow, with frequencies of the order of a few KHz. On the other hand, transverse motion
is dominated by the magnetic vector potential of each individual magnet represented by homogeneous polynomials.
This motion is generally much faster with frequencies of the order of a few MHz. Hence, the two motions
can be decoupled, at first approximation (e.g. the electric field component can be ignored). In addition, the magnetic fields can be considered static within a magnet and
transverse to particle motion, so that  the magnetic vector potential has only one component, the longitudinal
one. Following Maxwell equations and the theory of analytic functions, it can be shown that the vector
potential component for a certain magnet becomes~\cite{Rossbach}
\begin{equation}
A_z(x,y) = -B_0 r_0 {\mathfrak Re} \sum\limits_{n=0}^\infty
\frac{b_{n}(s) - i a_{n}(s)}{n+1} (\frac{x+iy}{r_0})^{n+1} \;\;,
\label{eq:vectpot}
\end{equation}
where $r_0$ is the reference radius which is chosen to be the
outermost conceivable deviation of the particles  and $B_0$ the main
 field of the magnet. 
The coefficients $b_n(s)$ and
$a_n(s)$ are named  the normal and skew multipoles, due to the associated magnetic field 
symmetry imposed by the homogeneous
polynomials. In this convention,  the dipole terms are represented
for $n=0$, the quadrupole for $n=1$, the sextupole for $n=2$, the octupole for $n=3$, etc.

Taking into account that the total momentum is much larger than the transverse 
(generally true for high energy accelerators), the square root of the Hamiltonian~\eqref{eq:SPRHamp} can be
expanded to leading order. Following some further approximations for large rings (imposing $x\ll\rho$) and applying additional canonical transformations, by rescaling the transverse momenta to the reference one and moving the periodic orbit to the origin~\cite{Ruth}, the Hamiltonian takes the form
\begin{equation}
{\cal {H}}  = \frac{p_x^2+p_y^2}{2(1+\delta)} - \frac{x(1+\delta)}{\rho(s)} - e\hat{A}_s \;\;,
\label{relhamexp}
\end{equation}
where $\delta = \frac{P_t-P_0}{P_0}$ is the relative momentum deviation of the particles' momentum $P_t$ with
respect to the reference momentum of the accelerator $P_0$ and $\rho$ the radius of curvature of the reference orbit. This $s$-dependent Hamiltonian has now 2 degrees of freedom plus the path length $s$ and for a ring it is periodic with period the circumference.

The Hamiltonian then
can be written in a more suitable form for dynamical analysis, by separating it to an integrable part 
and a non-integrable one:
\begin{equation}
{\cal H'} = {\cal H}_0 + \sum_{k_x,k_y} h_{k_x,k_y}(s) x^{k_x} y^{k_y}\;\;,
\label{relhamnonint}
\end{equation}
where  $h_{k_x,k_y}(s)$ are associated with the multipole coefficients and thereby are also periodic. The integrable 
Hamiltonian is derived by considering only dipole (uniform) fields with radius of curvature $\rho(s)$ and normal quadrupole (linear) magnetic fields,
with normalised gradient $K(s)$:
\begin{equation}
{\cal {H}}_0  = \frac{p_x^2+p_y^2}{2(1+\delta)} - \frac{x\delta}{\rho(s)} + \frac{x^2}{2\rho(s)^2} + \frac{K(s)}{2} (x^2 - y^2) \;\;.
\label{relhamnon}
\end{equation}
The  equations of motion are HillÕs equations (harmonic oscillators with periodic coefficients) and can be solved
following Floquet theory~\cite{CourantSnyder}. For analysing the non-integrable part,  
classical perturbation theory was employed already in the early days of the first synchrotrons ~\cite{Hagedornetal, Moseracc, 
Hagedorn, HagedornSchoch, Schoch}. 

\subsection{Accelerator maps and normal forms}

\begin{figure}[t!]
\begin{center}
\includegraphics*[angle=90,clip,width=12cm]{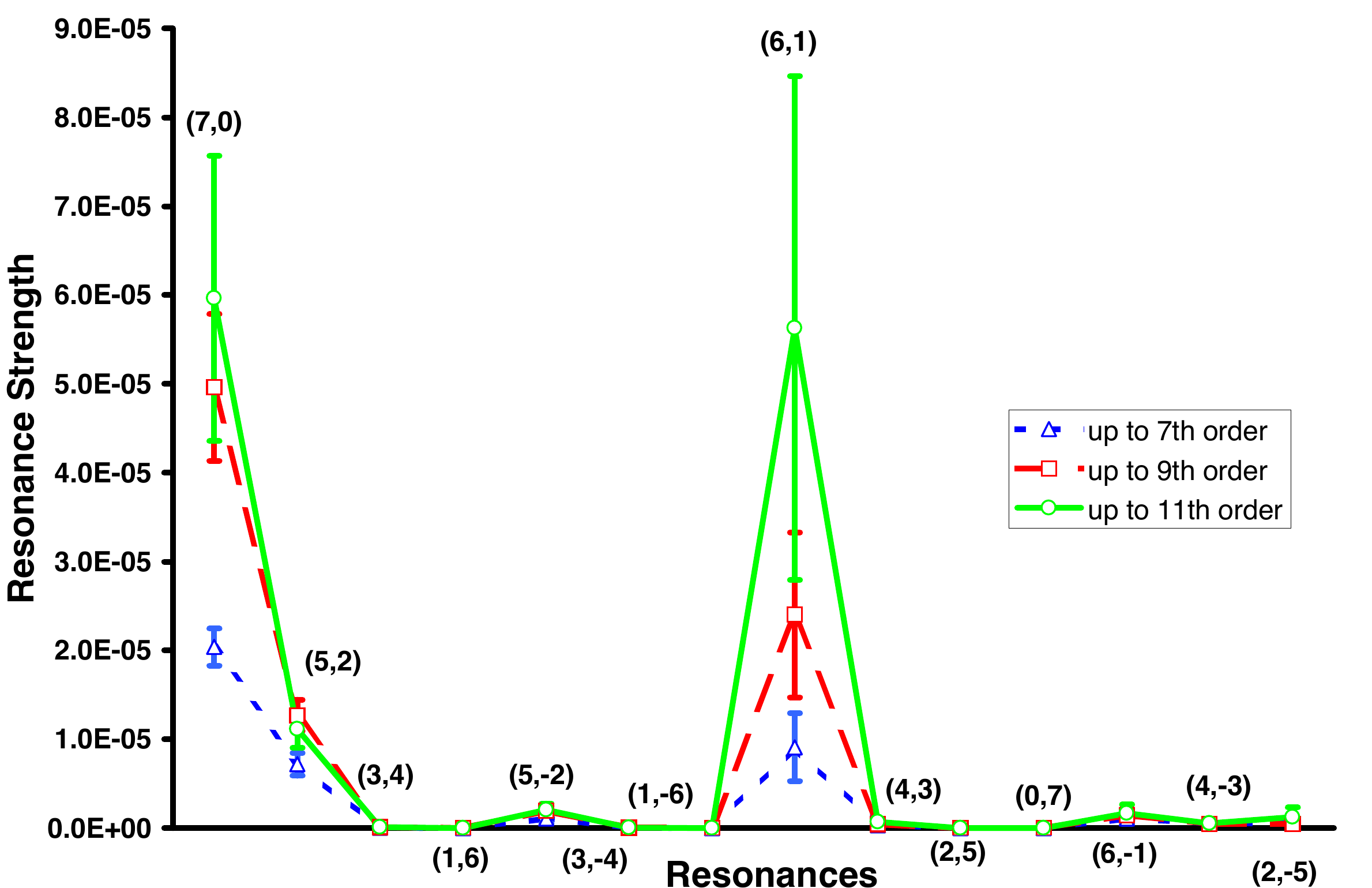}
\end{center}
\caption{\it Average value (circles) and standard deviation (error bars) of the 7th order
    resonance strengths, computed at an amplitude close to the minimum chaotic boundary,  
  over 60 random realisations of the multipole errors of an LHC magnet model. The three
    different curves represent resonance driving terms estimated by taking into account 
    three different orders of the generating function produced by a normal form analysis of a numerical one-turn map~\cite{GRR}.}
\label{ord7}
\end{figure}

Classical perturbation theory permits the estimation of the "resonance driving terms", the Fourier amplitudes associated to a particular resonance, issued by an expansion of the periodic perturbation of the Hamiltonian, in a Fourier series. In principle, the technique works for arbitrary order in the Taylor expansion of the perturbation, but the disentangling of variables becomes difficult even to 2nd order. This can be solved by the introduction of the Lie formalism, where the application of perturbation theory becomes completely algorithmic. In addition, the same formalism can be used for building and analysing a symplectic accelerator map~\cite{DragtFinn,Dragt79,Dragt82,Forestbook}. This  becomes even more appealing in the case of the accelerator rings, made out of thousands of elements.

 For a general Hamiltonian system 	
$H({\bf z}, t) $,  the HamiltonÕs equations  $\frac{d{\bf z}}{dt}=[H,{\bf z}] = :H:{\bf z}$ are represented by the usual Poisson bracket operator $:f: g = [f,g] =\sum_{i=1}^{N} \left( \frac{\partial f}{\partial p_i}
\frac{\partial g}{\partial q_i} - \frac{\partial f}{\partial q_i}
\frac{\partial g}{\partial p_i}\right)
 $. A formal solution  is
written as 
\begin{equation}
{\bf z}(t) = \sum\limits_{k=0}^{\infty} \frac{t^k{:H:}^k}{k!} {\bf
z}_0 = e^{t:H:}{\bf z}_0 \;\;, 
\label{Lie}
\end{equation}
with $\mathcal{M}=e^{:H:}$ a symplectic map in its usual exponential Lie operator representation.
The one-turn accelerator map is a polynomial of degree $m$ in the conjugate variables $z_1,\dots,z_n$, 
represented by the composition of the maps			 
$$\mathcal{M}= e^{:f_2:} \; e^{:f_3:} \; e^{:f_4:} \; \dots\;\;,$$	     
where the generators  $f_i$  are related to the Hamiltonians for each element.
Differential algebra tools can be used for computing efficiently the map~\cite{Berz}.
The construction of normal forms on the map consists of finding a symplectic transformation which
so that it becomes simpler. The generating function of this transformation can be represented by a polynomial
in the new variables, whose coefficients represent the "strength" of various resonances of the accelerator. The numerical application of these technique in the numerical map enable the evaluation of resonance strengths and can be used for dictating correction schemes~\cite{ForestBerzIrwin,BaToTuSe}.
An example of this normal form analysis for the LHC is shown in Fig.~\ref{ord7},
where a numerical tool was built in order to evaluate and represent graphically 
resonance strengths up to any order~\cite{GRR}. In this plot, the 7th order resonance strengths are
depicted,  issued by a normal form analysis of a numerically constructed 11th order map, 
for a given amplitude, close to the minimum chaotic boundary of this LHC model.  The 
points correspond to the average value of the resonance driving terms
over the 60 random realisations of the magnet errors and the error
bars are equal to one standard 
deviation. The three different lines represent 7th order resonance strengths
the computation of which is conducted up to three different orders in the generating function of the map (7th, 9th and
11th). With this method, the predominance of the \mbox{(7,0)} and  \mbox{(6,1)} resonances was revealed
including the important contribution
of the higher orders in the generating function, for accurately computing the resonance strengths.  
Although it was thought that the dynamics of the LHC at injection is largely 
determined by the random multi-pole errors in the super-conducting dipoles, which fill a large part of 
the ring, the method enabled to discover the importance of a few warm quadrupoles~\cite{GRR}.

\section{Chaos detection methods - the Frequency map analysis}
\label{FRM}

The accelerator Hamiltonian~\eqref{relhamnonint} is not bounded, so particles
inside the chaotic regions of the system will diffuse to infinity, or in
a real accelerator will be lost in the vacuum pipe. The focus of 
non-linear beam dynamics is to estimate and take correcting
measures in order to increase the area in real space where
particles survive after some time, the so-called dynamic aperture (DA).
This quantity (or particle survival rates) can be measured in 
a real accelerator~\cite{Bruning,Chaoetal,Willeke95}, or calculated through
numerical integration of the equations of motions (called Òparticle trackingÓ), 
with codes optimised for this task~\cite{ref:madx-ptc,sixtrack}. This brute-force approach, 
although providing a clear measure of accelerator performance, presents 
clear drawbacks. Firstly, the simulation  has to be carried out for very large rings 
with thousands of elements and for particles which need to survive several
millions (or practically infinite number) of turns. This can be numerically heavy and
it has to be repeated for several random realisations of the magnetic errors, in order 
to provide some statistical confidence about the robustness of the results. This can be 
partially solved by detecting the chaotic behaviour of the trajectory with
some fast indicators such as the Lyapunov exponents~\cite{Schmidt91, Giovannozi97},
the variance of unperturbed action~\cite{Chiretal, Tennyson88, Irwin} or 
Fokker-Planck-like diffusion coefficients~\cite{BruningPhD, SenEllison}. On the other hand,
 none of these approaches can solve the more serious problem which is the need
 of understanding globally the systems phase space structure.

The aforementioned  high-order perturbation theory has been extensively used in beam
physics in order to provide insight regarding the non-linear dynamics of particle beams.  However, 
it is not easy to correlate the computed resonance driving terms with
the extent of the chaotic region, or to put it simply, which are the resonances
that limit the DA in order to provide correcting measures.

A method that bridges this gap and can be used both as an early chaos
indicator but also for numerically estimating resonance strengths,
through tracking or measured data, is the Frequency Map Analysis (FMA).
 This  method has been
extensively used in celestial mechanics~\cite{Laskar88,Laskar90, LaskarRobutel}, galactic
dynamics~\cite{PapLas96, PapLas98}, atomic physics~\cite{Uzer}
Hamiltonian toy models~\cite{LaFrCe,Laskar93,DumasLaskar} and became 
a standard tool of beam dynamics analysis for a variety of accelerators, 
such as  hadron colliders~\cite{frmap,Yannis,PapZim99, 
PapZim02,Luo2012}, synchrotron light sources~\cite{LaskarRobin,LaskarPRL00,NaLa03,expfrmap03,expfrmap04}, 
high-intensity rings~\cite{YannisSNS,BartosikPS2}, B-factories~\cite{Liuzzo} and 
linear collider damping rings~\cite{Fanouria}.  The method relies on the high precision
calculation~\cite{ref:NAFF1} of 
the associated frequencies  of motion (or "tunes" in the accelerator jargon), 
which are supposed to be invariant in the case of quasi-periodic motion,
as stated by the KAM theory. In this respect, the variation of the frequencies
over time~\cite{Laskar93,DumasLaskar,frmap} can provide an excellent early
stability indicator that has the advantage of connecting resonant structure
with diffusion of chaotic trajectories. 

The first step is to derive through the Numerical Analysis of Fundamental Frequecies (NAFF) algorithm 
\cite{Laskar88,Laskar90} or variants of this code~\cite{SUSSIX}, a
quasi-periodic approximation, truncated to order $N$,
\begin{equation}
 f'_j(t) = \sum^N_{k=1} a_{j,k} e^{i\omega_{jk}t} \; ,
\end{equation}
with $f'_j(t), a_{j,k}\in \mathbb{C}$ and $j=1,\dots,n$,
of a complex function $f_j(t)= q_j(t) + i p_j(t)$, formed by a pair of
conjugate variables of a general $n$-degrees 
of freedom Hamiltonian system, which are determined by usual numerical
integration or experimentally measured, for a finite time span $t=\tau$. The next step is to
retain from the quasi-periodic
approximation the fundamental frequencies of motion, corresponding most of the times to the frequency of the dominant Fourier component $\omega_{j1}$, for each degree of freedom. In this respect, the frequency vector $\frac{\boldsymbol \omega}{2\pi} = {\boldsymbol \nu} =  (\nu_1,
\nu_2, \dots, \nu_n)$ can be constructed, which, up to numerical accuracy
\cite{ref:NAFF1}, parameterizes the KAM tori in the stable regions of a
non-degenerate Hamiltonian system. Then, the construction of the
frequency map can take place~\cite{LaFrCe, Laskar93, DumasLaskar,
LaskarRobin}, by repeating the procedure for
a set of initial conditions which are transversal to the orbits of
interest. As an example, we may keep all the momenta ${\boldsymbol  p}$
 constant, and explore the positions ${\boldsymbol q}$ to
produce the map ${\mathcal F_\tau}$:
\begin{equation}
{\mathcal F_\tau}\;:
\begin{matrix} 
{\mathbb R}^{n}&\longrightarrow &{\mathbb R}^{n} \\
{\boldsymbol q}|_{{\boldsymbol p}={\boldsymbol p_0}}&\longrightarrow
&{\boldsymbol \nu} \;.
\end{matrix}
\end{equation}
The dynamics of the system is then analysed by studying the regularity
of this map on frequency or initial condition space. In addition, each initial condition 
can be associated with a
diffusion indicator, by computing the frequency vector for two equal and
successive time spans, which correspond to half of the total
integration time $\tau$. The amplitude of the diffusion vector,
\begin{equation}
{\boldsymbol D}|_{t=\tau} = {\boldsymbol
  \nu}|_{t\in(0,\tau/2]}-{\boldsymbol \nu}|_{t\in(\tau/2,\tau]} \;,
  \label{Diffvec}
\end{equation}
 can be used for characterising the instability
of each orbit.
Through this representation  the
traces of  the resonances can be viewed in the physical space, as well, and set a 
threshold for the minimum DA. Moreover,  a diffusion
quality factor defined as the average of the local diffusion
coefficient to the initial amplitude of each orbit, over a domain $R$
of the phase space:
\begin{equation}
D_{QF} = \big\langle \; \frac{{|{\boldsymbol
      D}|}}{|\boldsymbol q_0|} \;  \big\rangle_R \;.
      \label{DiffQF}
\end{equation}
This quantity can be used for the comparison of different designs  
and the optimisation of the correction schemes proposed.

\section{Application of frequency map analysis in accelerator models}
\label{accelFMA}

\subsection{Frequency maps for the LHC}

The long term stability of the beam is the major concern for the
design of a hadron collider, as the LHC. Especially during
long injection period of more than $10^7$ turns needed to fill the LHC
with more than 2800 bunches per beam, in its nominal configuration, particle trajectories are perturbed
strongly by non-linear magnet fields, mainly attributed to the
multipole errors of the super-conducting magnets.
\begin{figure*}[ht]
\begin{center}
    \includegraphics*[height=6.4cm,width=7.1cm]{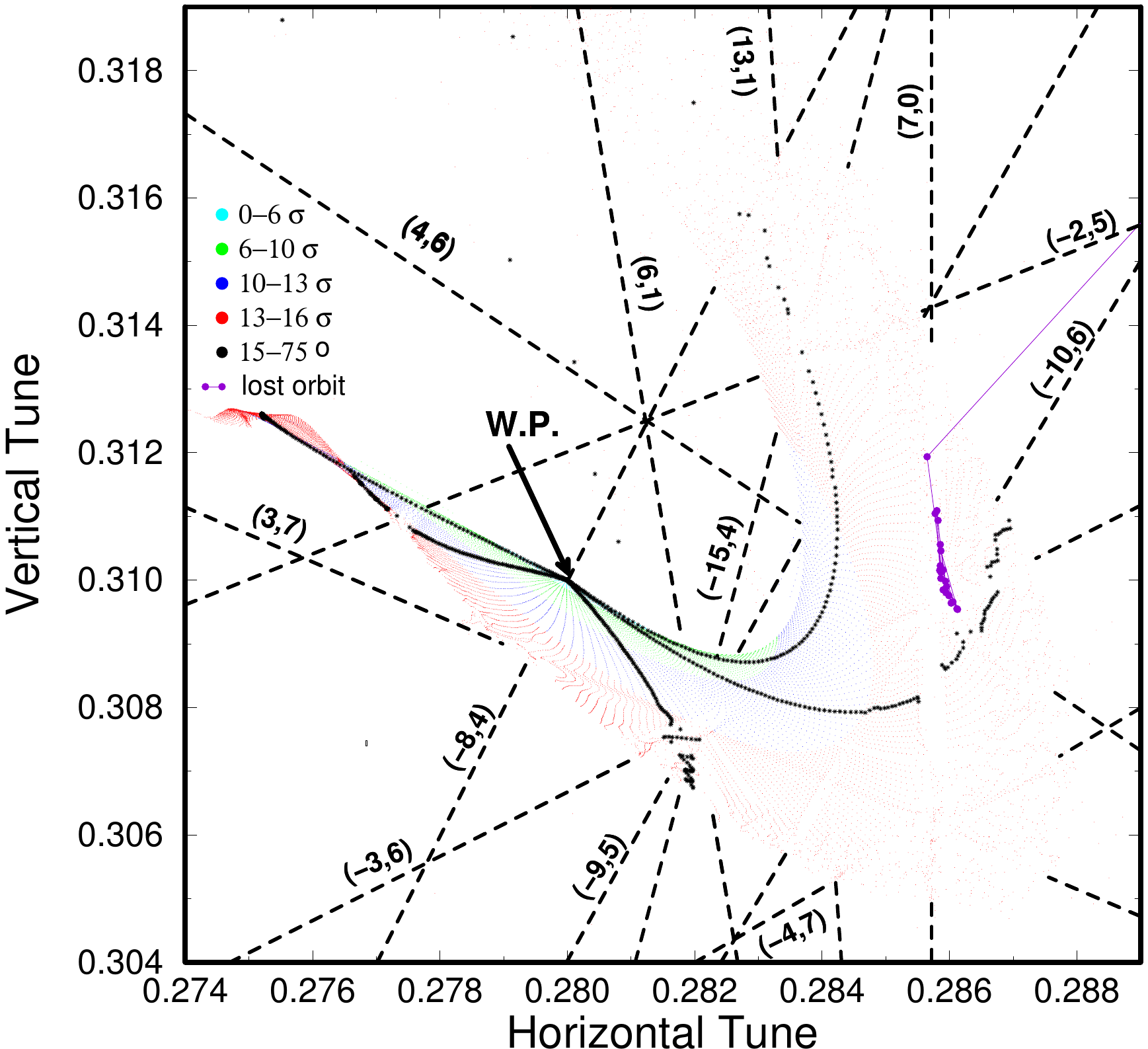}\hskip 1.cm
    \includegraphics*[height=6.4cm,width=7.1cm]{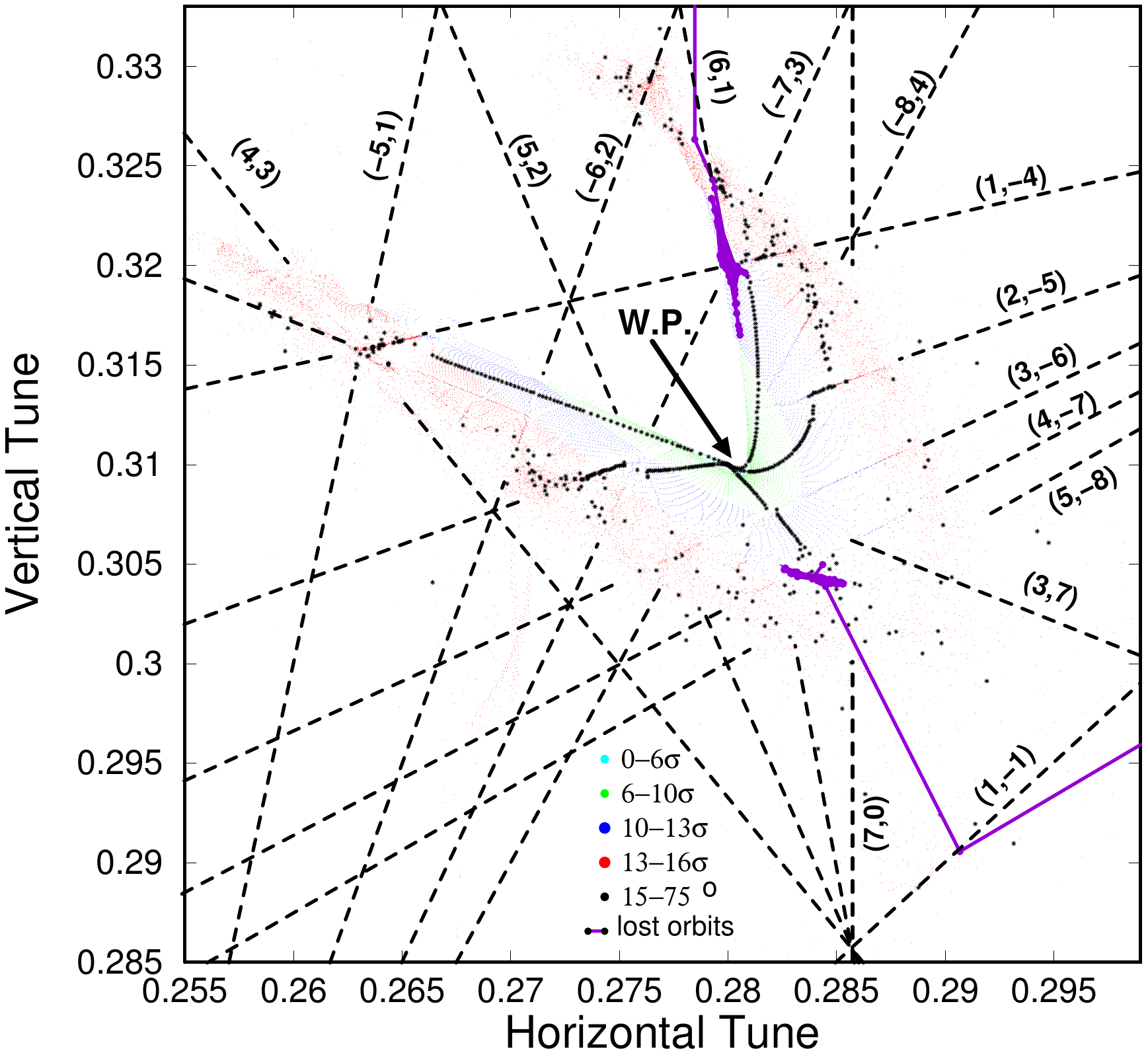}  \ \vskip 0.3cm
    \includegraphics*[height=6.4cm,width=7.1cm]{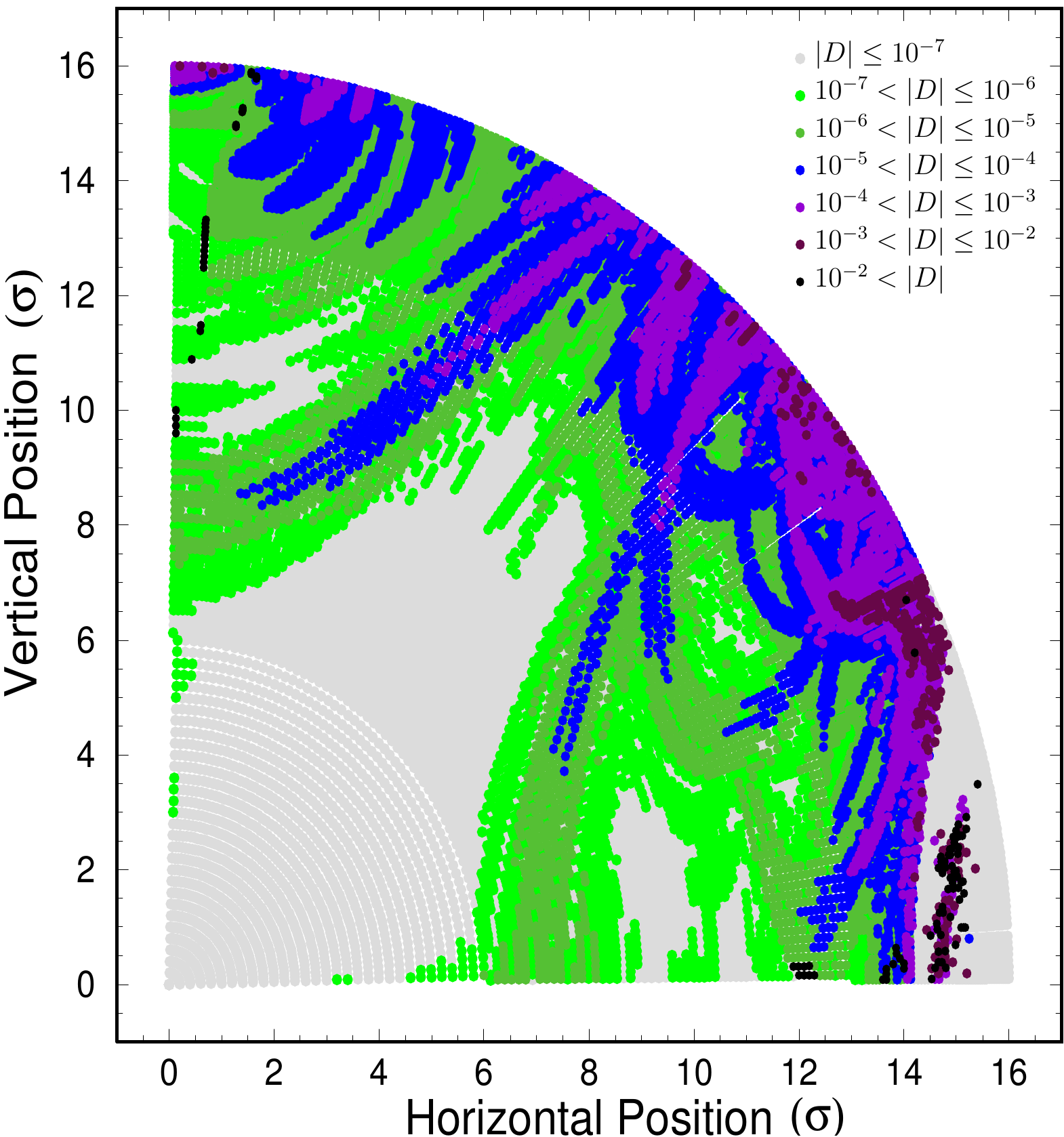}\hskip 1cm
    \includegraphics*[height=6.4cm,width=7.1cm]{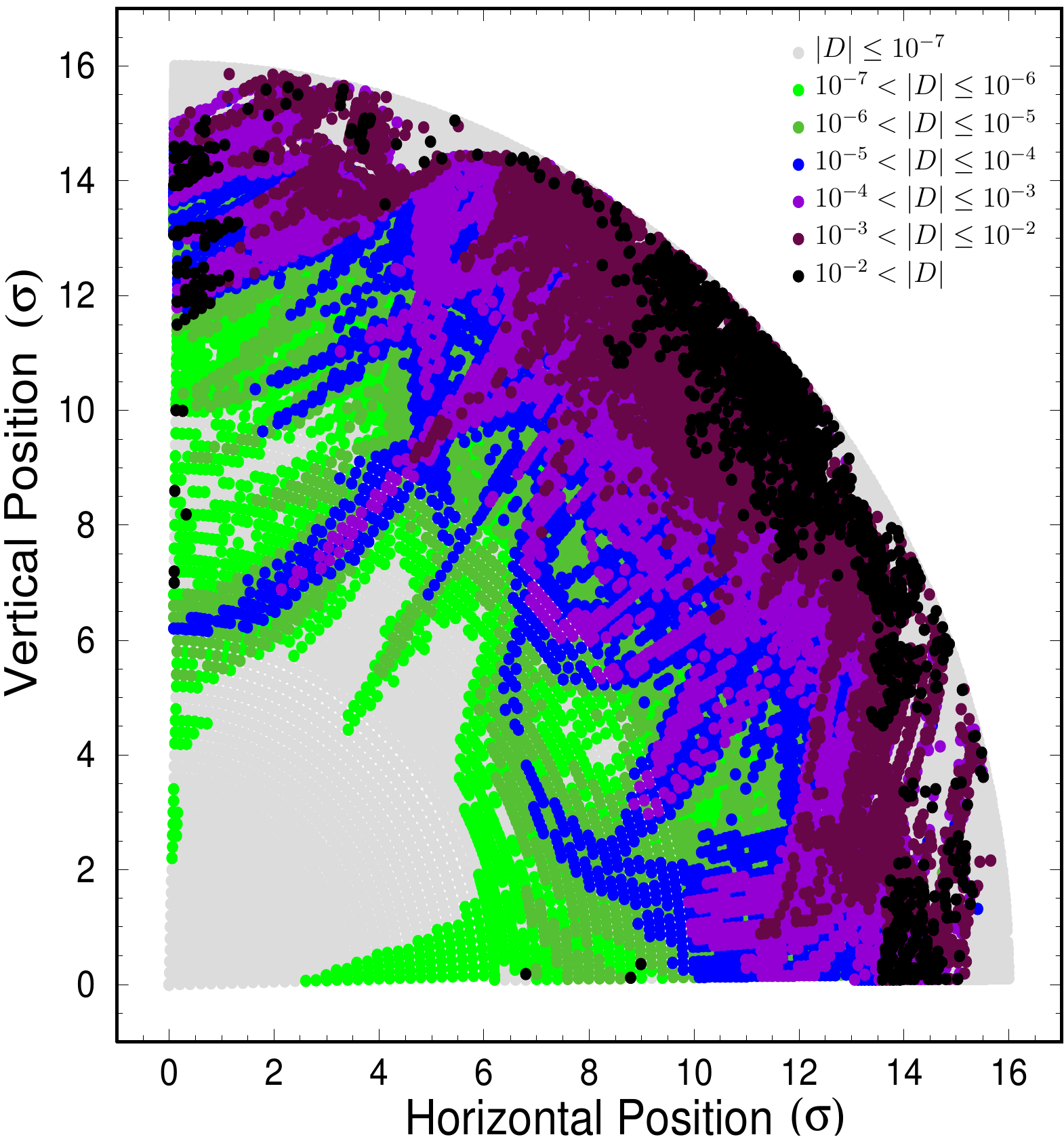} 
\end{center}
\caption{Frequency (top) and diffusion (bottom) maps for the LHC  target
  error table without (left) and with (right) the high
  $a_4$ value on the dipoles.~\cite{frmap}}
\label{frmap5}
\end{figure*} 
In order to reach the target DA of 12 rms beam sizes $\sigma_x=\sigma_y=\sigma$,  for the LHC injection
optics, giving a necessary safety factor of 2 with respect to the
physical aperture restriction of the beam  at $6~\sigma$ (location of collimators
protecting the super-conducting magnets), 
a target magnetic multi-pole error table  was proposed. A frequency map 
for this error table and for the nominal ÒtunesÓ ($Q_x=64.28$, $Q_y=59.31$) is shown
in the top left part of  Fig.~\ref{frmap5}~\cite{frmap}.  This specific machine gives an average
DA  of around $13\sigma$ and a minimum of $12\sigma$, values which are close to the average 
and minimum DA over all the 60 random realisation of the magnet errors. Each point in the 
frequency space corresponds to a particular on-momentum orbit tracked over 1000 turns. 
The different colours in the map correspond to orbits with
different initial position amplitudes (from $0-16\sigma$)
and the black dots label initial conditions with   
different amplitude ratios (from $15^\circ$ to $75^\circ$). The orderly spaced
points correspond to regular orbits whereas the dispersed points to
chaotic ones. This plot is a snapshot of the so called Arnold web~\cite{Arnold}, the
 network of resonances $a\nu_x+b\nu_y+c=0$, which appear as
distortions of the map (empty and filled lines) and can be easily
identified. For example, the importance of three 7th order
resonances ($(a,b) = (7,0), (6,-1)\,\text{and}\,(-2,5)$) is put in evidence. 
Especially the crossings of the resonant lines are ``hot spots'', from which
particles can easily diffuse: as an example, the evolution of
the frequency vector of an orbit starting close to the crossing of the $(7,0)$
with the $(-3,6)$ and $(4,6)$ resonances is shown (purple dots).  The orbit diffuses along
the unstable manifold of the 7th order resonance and
is lost after a few thousand turns. This is a clear demonstration of
the importance of this resonance with respect to the DA of this model.

One of the main issues in the specification of the LHC injection
optics, is the correction of the systematic part of the lowest order
multipole errors of the super-conducting dipoles, which limit the
DA~\cite{Koutch}. This is usually done by magnetic coils (``spool
pieces") placed at the ends of the dipoles. In the case of  more realistic  error 
table with increased normal and skew octupoles, there was an important loss of the 
DA~\cite{skew} with respect to the target error table. A
frequency map for the same random ``seed" as for the previous case with the
increased skew octupole error in the dipoles is shown in the top right plot of
Fig.~\ref{frmap5}. The frequency maps now looks much more distorted. The
most remarkable feature concerning 
the system's dynamics is the huge increase of the tune variation with amplitude, to the
point that particles are diffusing towards the 
$(1,-1)$-resonance, in the right bottom corner of the map. On the other hand, particles
close to horizontal motion at the top of the map are approaching the
$(0,3)$ resonance and the ones close to vertical motion the $(4,0)$. 
This finding has been confirmed with Normal Form analysis~\cite{skew}. The dynamic
aperture could be recovered by tuning the skew octupole spool pieces
such as to cancel the $(1,-1)$ resonance~\cite{skew}.

The global dynamics of these two cases can be also explored in the physical
space of the system by mapping each initial condition with the diffusion vector~\eqref{Diffvec}, 
the amplitude of which can be used for characterising the instability
of each orbit. In Figs.~\ref{frmap5} (bottom),  the points in the
configuration space are plotted using a different colour coding corresponding to
different diffusion indicators in logarithmic scale: from grey for
stable ($|{\boldsymbol D}|\le 10^{-7}$) to black for strongly 
chaotic particles ($|{\boldsymbol D}|>10^{-2}$), that actually are lost
within that short integration time. Through this
representation, the traces of 
the resonances in the physical space are clearly visible, and thereby a 
threshold for the minimum DA can be set.

\begin{figure}[t]
\begin{center}
\rotatebox{0}{\scalebox{0.44}{{\includegraphics*{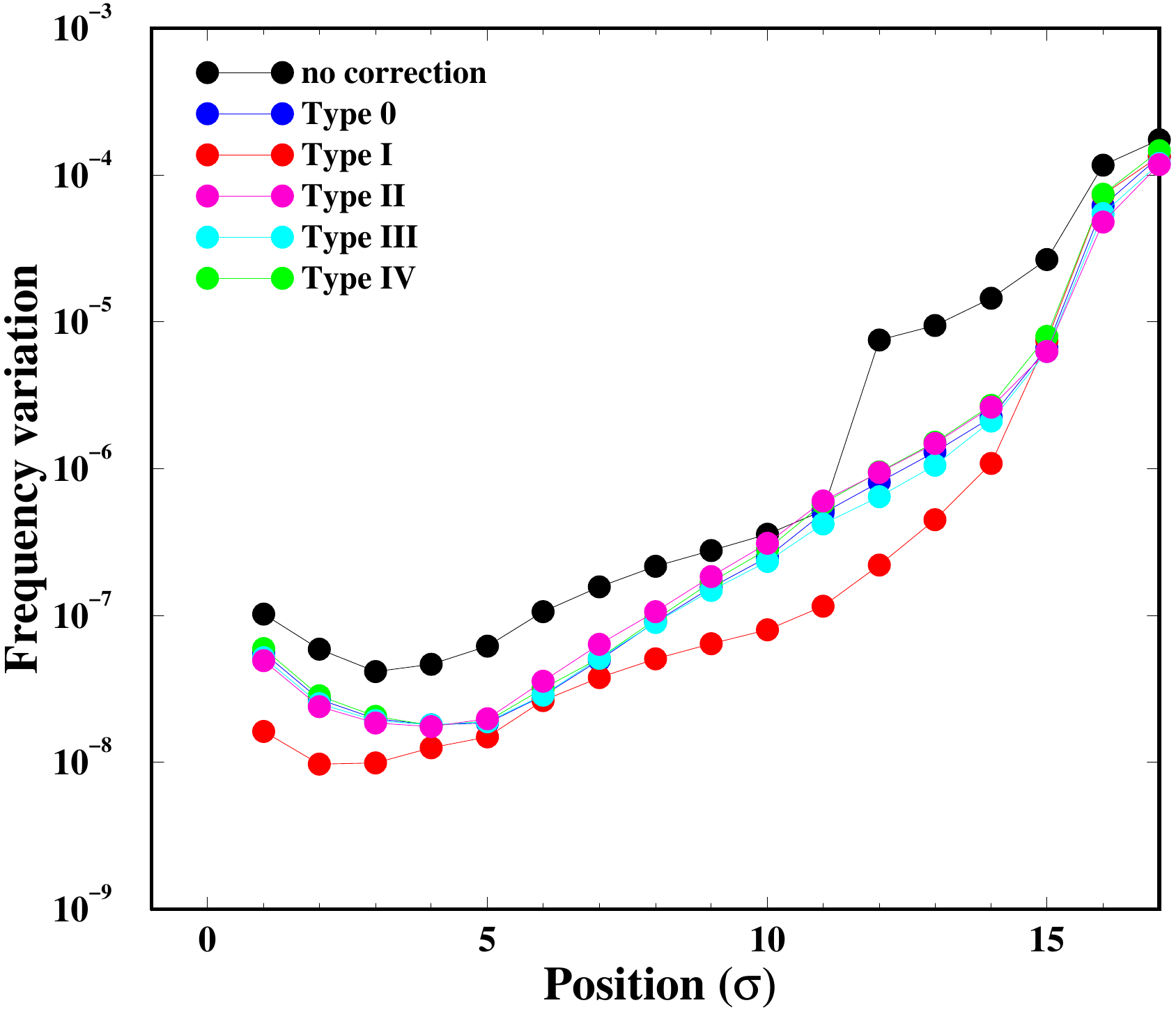}}}}\hskip .5cm
\rotatebox{0}{\scalebox{0.44}{{\includegraphics*{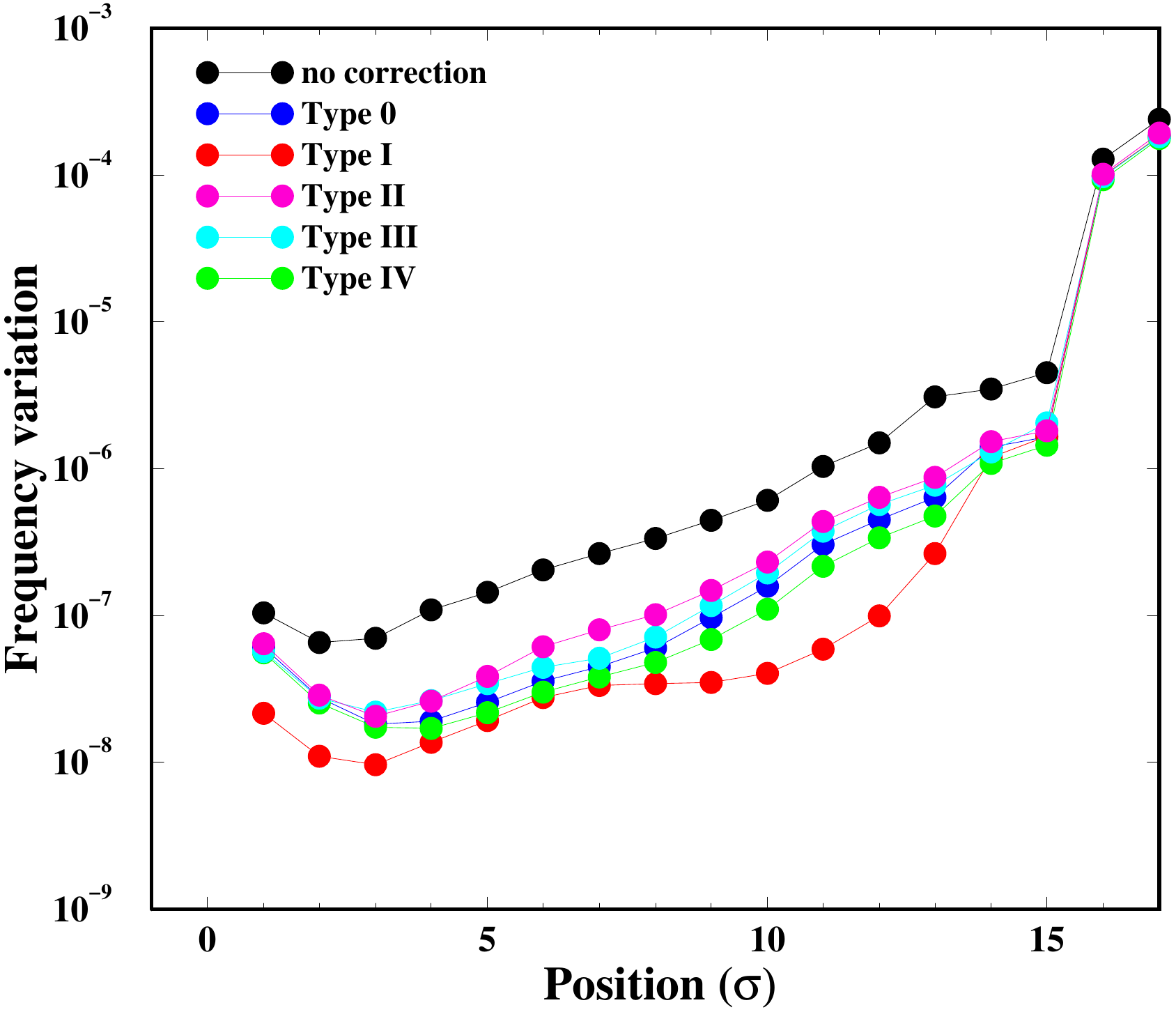}}}}
 \caption{Evolution of the frequency variation averaged over all
directions, with the particles' amplitude (in $\sigma$) for $\delta
p/p=7\times10^{-4}$,  and for two different tunes: $(Q_x,Qy)=(0.28,0.31)$  (left) and $(Q_x,Qy)=(0.21,0.24)$ (right)~\cite{Yannis}.}
\label{diffu}
\end{center}
\end{figure}

The diffusion
quality factor~\eqref{DiffQF}  is a very efficient global chaos indicator for comparing different designs  
and optimising the correction schemes proposed~\cite{Yannis,BartosikPS2,ALS2012}. For
example, for the normal octupole $b_4$ and decapole $b_5$ correctors in the LHC, 
five schemes where proposed, regarding the positioning and the amount of
the correctors~\cite{Yannis}.
Frequency maps were produced for all the correction
cases and two working points.
In Fig.~\ref{diffu}, the diffusion quality factor averaged over the angles 
is plotted in logarithmic scale  versus the amplitude, for both working points,
 for all correction schemes and for the non-zero momentum deviation. These plots confirmed
 that all the correction schemes are quite similar and indeed necessary, following the comparison
 with the diffusion quality factor with no correction
 (black dots). They also indicated that the nominal (but most expensive) solution of including
 correctors in all the super-conducting dipole (blue dots)
was not performing better than the one with correctors in every second dipole (red dots), which
actually presented a slightly better diffusion quality factor. Based on this study, the baseline
correction for the LHC was to have correctors in every second dipole, which was also
a cost effective solution. Finally, this study demonstrated that the diffusion quality factor  
is correlated with other global chaos indicators, such as the resonances driving terms norm 
and the dynamic aperture~\cite{Yannis}.

\subsection{Working point choice through frequency maps}

\begin{figure}[htb]
\centering
\includegraphics*[trim= 10 230 10 260, clip,width=150mm]{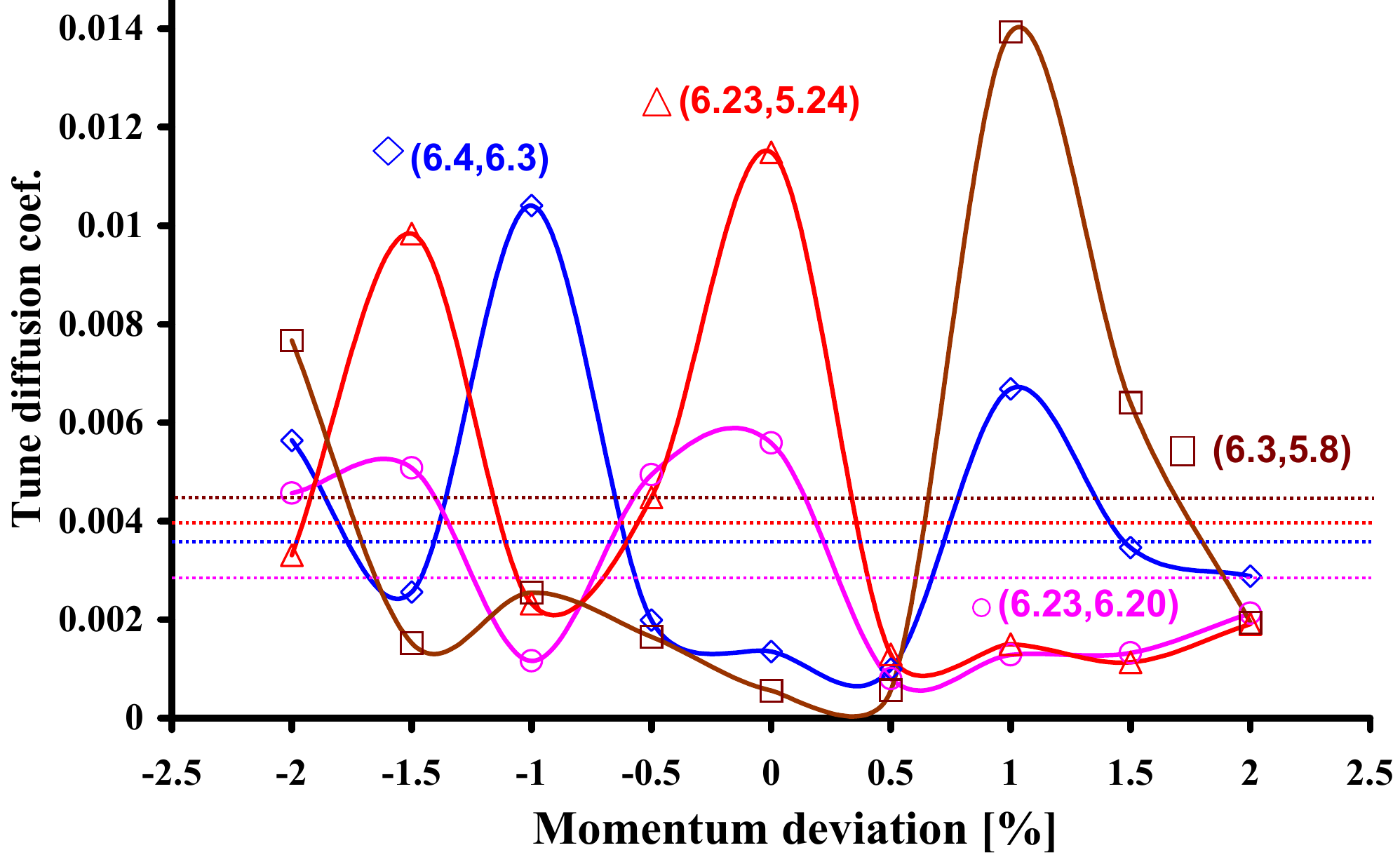}
\caption{Tune diffusion coefficients for four SNS working points versus the
different momentum deviations.~\cite{YannisSNS}
}
\label{fig:tunecomp}
\vspace{-15pt}
\end{figure}

Frequency map analysis is naturally very powerful for choosing the
best working point of an accelerator. An example is given in 
Fig.~\ref{fig:tunecomp}, where the value of the tune diffusion
coefficient  is plotted versus the momentum deviation
$\delta p/p$, for all working points of the Spallation Neutron Source accumulator
ring~\cite{YannisSNS},  currently in operation in Oak-Ridge
National Laboratory and holding the beam power world record.  The single-particle dynamics of this ring is dominated by edge effects
in the magnets, and especially the quadrupole
fringe fields, which are octupole-like~\cite{Forestbook}. Four working points were selected and compared
corresponding to the different curves of Fig.~\ref{fig:tunecomp}.
 The peak values on the diffusion indicators,
for all working points correspond to areas of the phase space that are
perturbed due to 4th order resonances, showing  the
destructive effect of quadrupole fringe fields. The dotted lines on the
plots represent the average values of the diffusion indicators for all
tracked momentum deviations. It is clear that (6.23,6.20) is the best
choice, followed by (6.4,6.3). Their performance can be further
improved by using the available multi-pole correctors~\cite{YannisSNS},
for correcting the normal and skew 3rd order resonances, in the case
of (6.4,6.3), and the 4th order normal resonances in the case of
(6.23,6.20).  The other two  working points have the disadvantage of crossing
major resonances, which are very difficult to correct. Based on this
study, the nominal working point of the SNS ring was chosen to be
(6.23,6.20) and was successfully used in commissioning and operation until today.

Similar working point optimisation was employed for different type of rings such as SuperB, a lepton
collider designed at INFN-LNF~\cite{Liuzzo}, or the PS2 ring~\cite{BartosikPS2}, an upgrade project of the current CERN Proton Synchrotron.

\subsection{Chaotic behaviour due to the long range beam-beam interaction}

In a colliding-beam storage ring, one of the largest perturbations affecting the motion of
beam particles is the collision with the opposing beam. This interaction occurs, unavoidably,
in the form of head-on collisions between bunches of the two beams at designated interaction
points. Hadron colliders employ long trains of closely spaced bunches, and individual bunches
encounter many others of the opposing beam at various long-range collision points, where the
beams are not fully separated into two disjunct
beam pipes. In general, the effect of the long-range collisions depends on the
ratio of the beam separation to the local rms beam size, and on the total number of long-range
collision points. On either
side of the two LHC main collision points, a beam encounters about 15 long-range collisions
with an approximate average separation between the closed orbits of the two beams of 9.5 rms
beam sizes.

\begin{figure}[t]
\begin{center}
{\includegraphics[height=8cm,width=8cm]{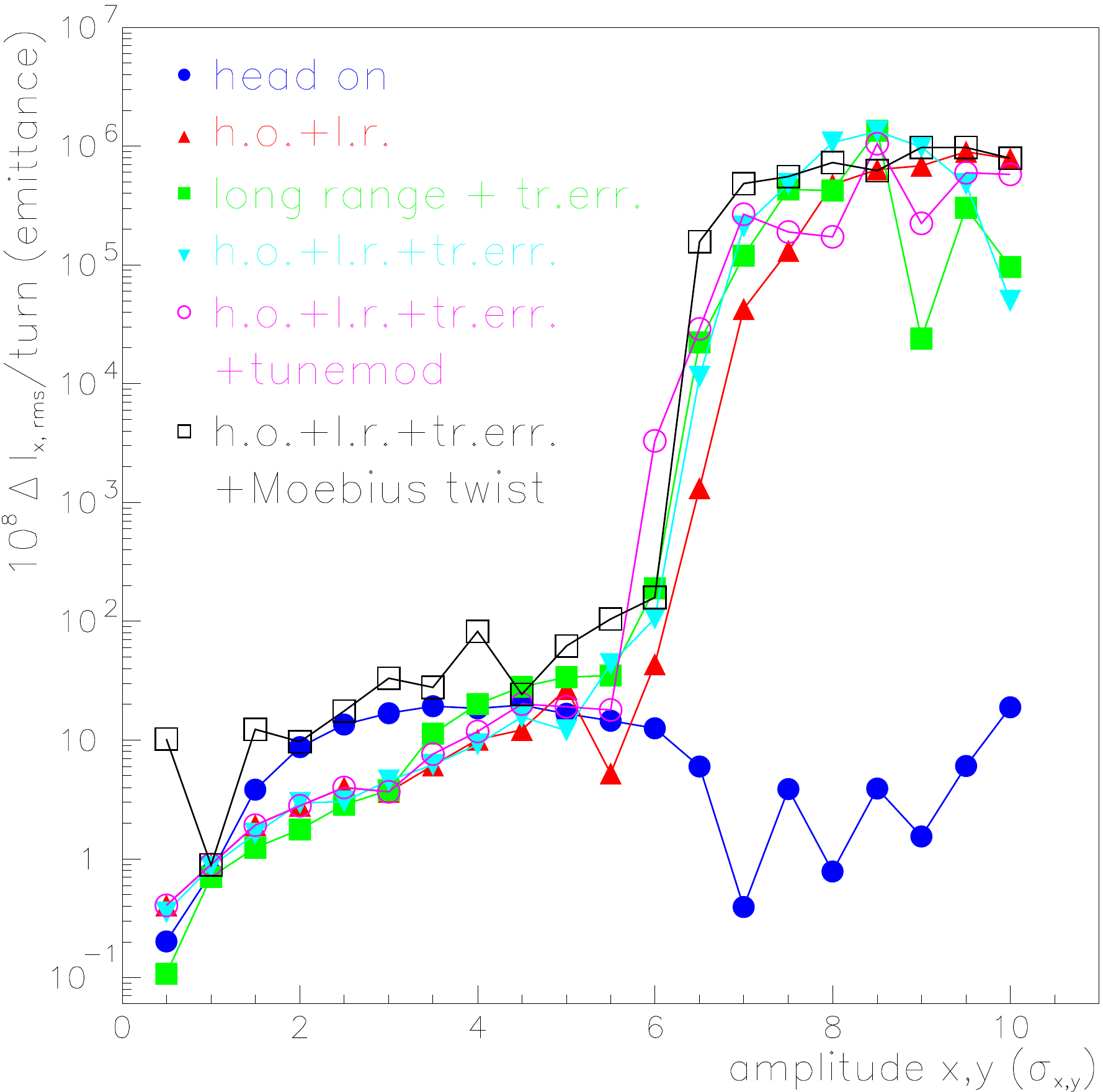}}
\end{center}
\caption{The change of action variance per turn
as a function of the starting amplitude. Compared are
the cases: head-on collisions only (dark blue), 
head-on and long-range collisions (red),
long-range collisions plus triplet field errors (green)
both types of collisions plus  triplet
field errors (light blue), the additional effect of a tune
modulation at the synchrotron frequency (22 Hz)
of amplitude $10^{-4}$ (pink), the additional effect
of a M\"{o}bius twist (black)~\cite{PapZim99}.
}
\label{diffam}
\end{figure}

Simulations predict that the long-range collisions in hadron colliders give rise to a well
defined border of stability~\cite{Irwin,PapZim99}. 
 Following standard ideas popularised by Chirikov 
and his collaborators \cite{Chirikov,Chiretal}, a diffusion
coefficient can be estimated by calculating the variance of the
unperturbed actions for a large number of turns.
Fig.~\ref{diffam} displays the change 
of the action variance, in terms of beam sizes, as computed 
by beam-beam simulations which consider
the particle motion in a 4-dimensional transverse phase space for a model with 2 interaction
points  and parameters similar to those of the LHC~\cite{PapZim99}. The stability border is insensitive
to the presence of the head-on collision (filled circles with dark blue curve), and only marginally
affected by the nonlinear field errors in the final-triplet quadrupoles (squares with green curve)
or by a small additional tune ripple (empty circles with pink curve). This ``diffusive aperture'' with
long-range collisions is equally insensitive to transverse closed-orbit offsets between the two
beams at the head-on collision points~\cite{PapZim99}.

\begin{figure*}
\begin{center}
{\includegraphics*[trim= 30 50 90 320, clip,height=7cm,width=7.5cm]{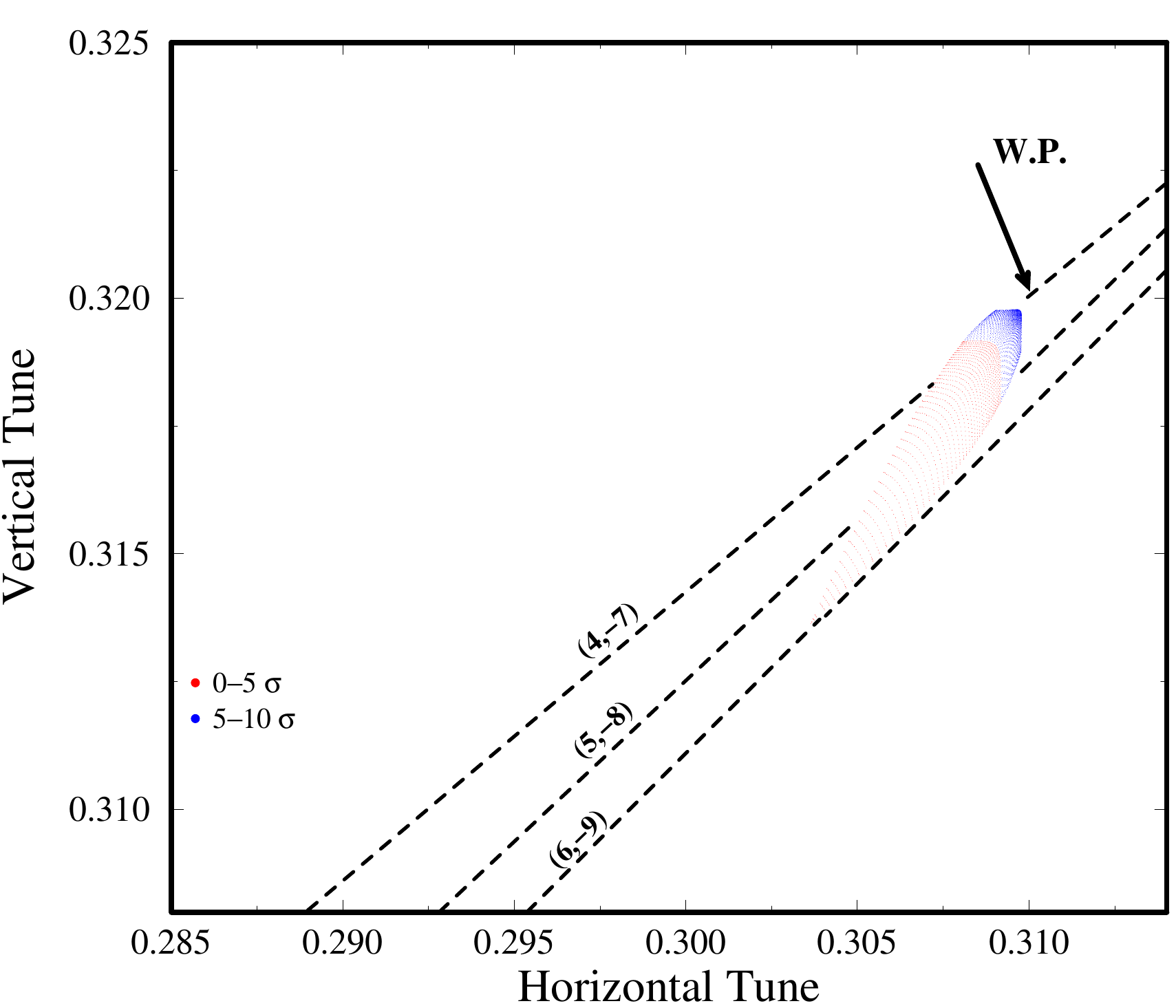}}\hskip 1.2cm
{\includegraphics*[trim= 30 50 90 320, clip,height=7cm,width=7.5cm]{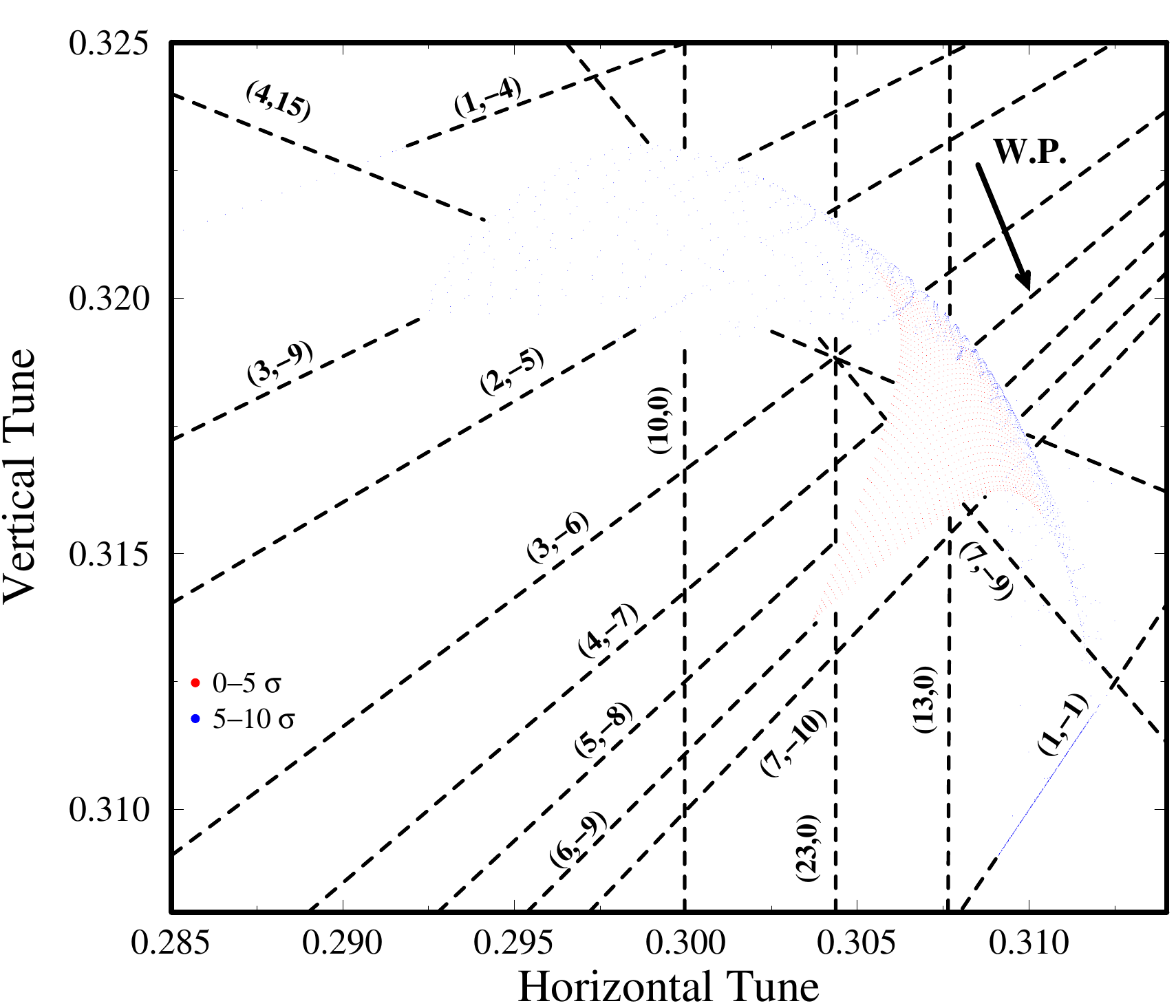}}\\
{\includegraphics*[height=6.8cm,width=7.5cm]{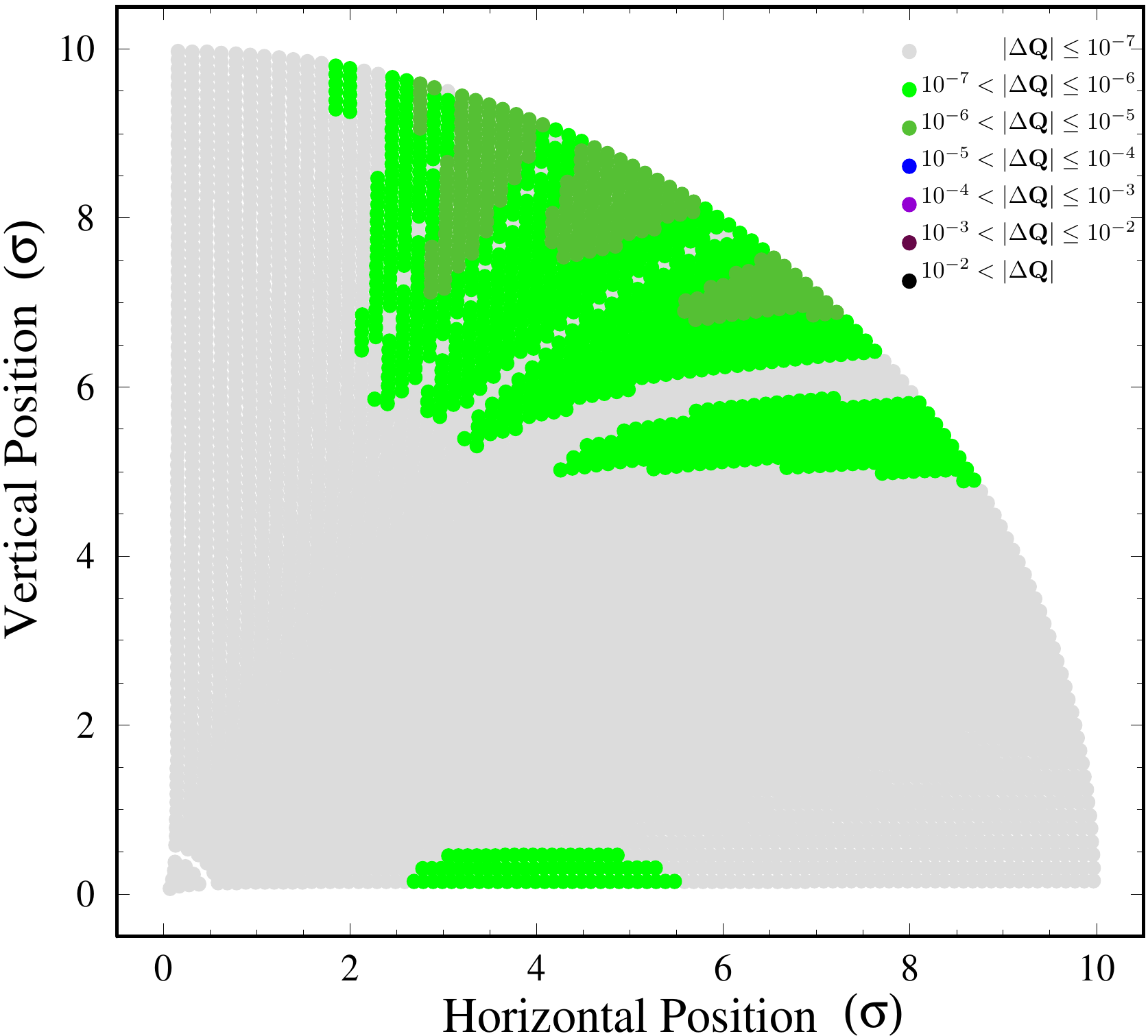}}
\hskip 1.cm
{\includegraphics*[height=6.8cm,width=7.5cm]{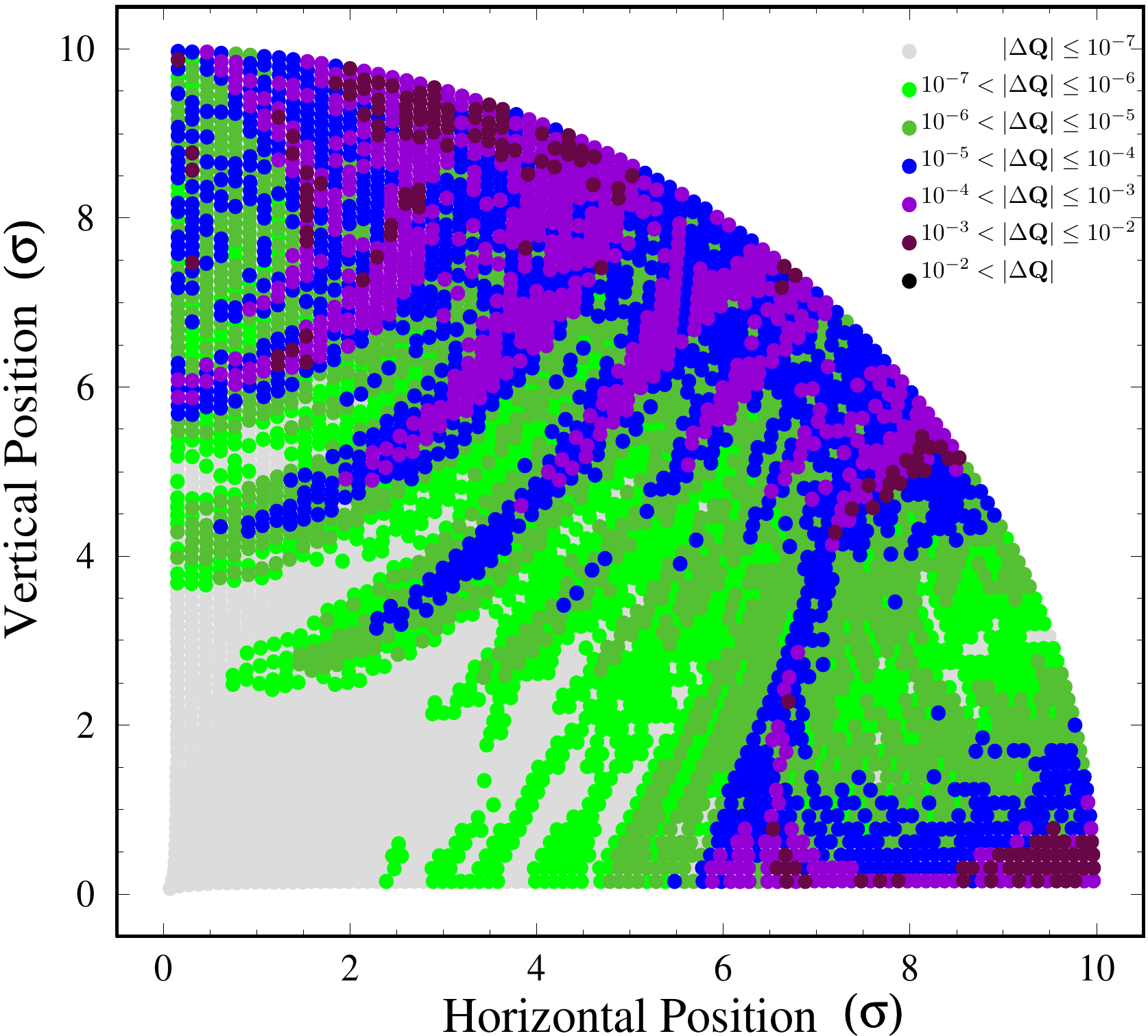}}
\end{center}
\caption{Frequency (top) and diffusion (bottom) maps for head-on
collisions only (left)  and including long-range collisions, as well (right)~\cite{PapZim99}.
} 
\label{footprint}
\end{figure*}

The top part of figure~\ref{footprint} presents frequency maps obtained by tracking single particles
over 1000 turns under the influence of beam-beam effects. Red dots
represent particles with initial transverse amplitudes up to 
5~$\sigma_{x,y}$, whereas blue dots show results for 
initial amplitudes up to 10~$\sigma_{x,y}$.
The dramatic effect of the long-range collisions is revealed through
the comparison of the maps obtained with (right) and without (left) the long-range effects. 
Up to initial particle amplitudes of around 6~$\sigma_{x,y}$, the effect of
the head-on collisions dominates. Then, the long-range effect
takes over and the frequency map flips, as the tune shift with
amplitude changes direction. This non-monotonic dependence of the tune
with respect to the amplitude is potentially dangerous for the
stability of particles beyond this limit~\cite{Laskar93,ref:NAFF2}.
The conclusions of the previous paragraph regarding the dominant
destabilising role of the long-range collisions are also confirmed
in the diffusion maps at the bottom plots of Fig.~\ref{footprint}.

\begin{figure}
\begin{center}
{\includegraphics*[height=7.5cm,width=7.5cm]{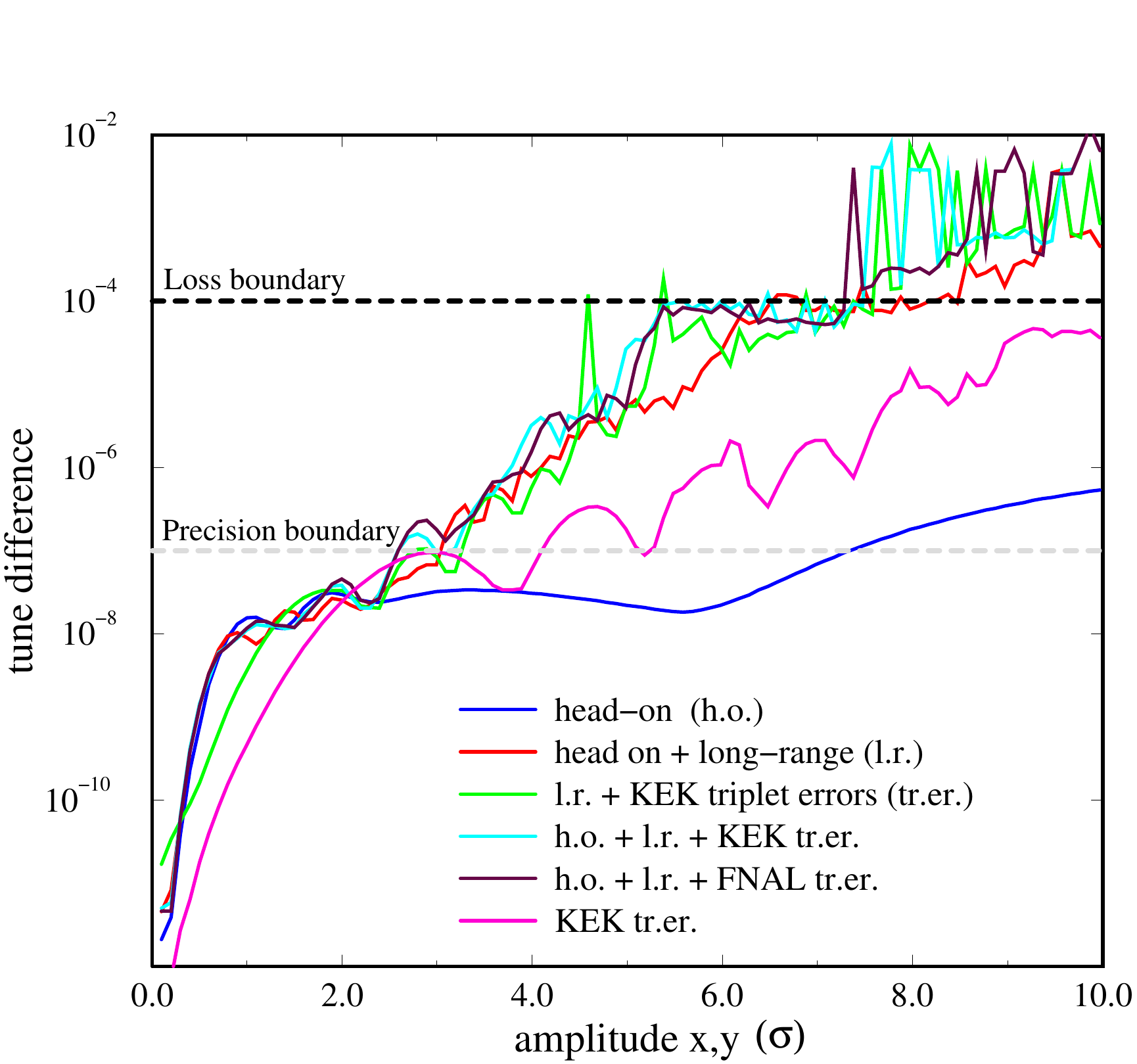}} \hskip 1cm
\end{center}
\caption{The change of frequency per 500 turns
averaged over all initial amplitude ratios $x-y$ as a function of the starting
amplitude. Compared are the cases: head-on collisions only; 
head-on and long-range collisions;
long-range collisions plus  triplet field errors;
both types of collisions plus  triplet
field errors; both types of collisions plus  triplet
field errors; only  triplet errors.
}
\label{df}
\end{figure}

In Fig.~\ref{df}, the diffusion quality factor versus the amplitude,
averaged over all initial amplitude ratios is plotted, 
for different combinations of beam-beam and  triplet nonlinearities (quadrupole magnets focusing
the beam at the collision point). 
There are two thresholds marking the precision boundary and a  particle loss boundary for tune changes bigger than
$10^{-4}$. Considering linear frequency diffusion over time, this corresponds to one unit in frequency within $10^7$ turns, which
certainly induces particle loss. For all the cases where long-range collisions and triplet
field errors are included, the loss boundary is located at the same
point, around $5.5 \sigma_{x,y}$. For the case where the triplet field
errors are not added to the beam-beam effect, the threshold is reached a
little further, around $6\sigma_{x,y}$. The case with only 
triplet errors is clearly more stable, but indeed there is still
a visible effect
for larger initial amplitudes. No effect whatsoever can be observed
for the case with only the head-on effect included, where the tune
variation is very close to the precision limit of the method.

\subsection{Dynamics of the CLIC Pre-damping rings}

The main limitation of the DA in the low emittance lattices comes from the non-linear effects induced 
by the strong sextupole magnet strengths, which are introduced for the  correction of the tune change with momentum (chromaticity). 
Following first order perturbation theory~\cite{ref:nonlinearsls}, the strength of a resonance $n_x Q_x+n_y Q_y=p$ of order $n$, with $|n_x|+|n_y|=n$ 
the order of the resonance and $p$ any integer, vanishes within an ensemble of $N_c$ cells, if the resonance amplification factor is~\cite{ref:ResFreeLat}
\begin{equation}
 \label{eq:resonancecancellation}
  \left| \sum_{\mathrm{p=0}}^{\mathrm{N_c-1}} e^{\mathrm{ip(n_x\mu_{x,c}+n_y\mu_{y,c})}} \right|=
  \sqrt{\frac{1-\cos[N_c(n_x\mu_{x\mathrm{,c}}+n_y\mu_{y\mathrm{,c}})]}{1-\cos (n_x \mu_{x\mathrm{,c}}+n_y \mu_{y\mathrm{,c}})}}=0\;\;,
\end{equation}
with $\mu_{x,y}$ the horizontal and vertical phase advance of the cell. Note that the tune is just the total phase advance for one turn.
The previous condition is achieved when $N_c(n_x\mu_{x\mathrm{,c}}+n_y\mu_{y\mathrm{,c}})=2k\pi$, provided the denominator of 
Eq.~\eqref{eq:resonancecancellation} is non zero, i.e.: $n_x\mu_{x\mathrm{,c}}+n_y\mu_{y\mathrm{,c}}\neq 2k'\pi$, with $k$ and $k'$ 
any integers. 
From this, a part of a circular accelerator will not contribute to the excitation of any non-linear resonances, except
of those defined by $\nu_x \mu_x +\nu_y \mu_y = 2 k_3 \pi$, if the phase
advances per cell satisfy the conditions: $N_c \mu_x = 2k_1\pi$ and $N_c \mu_y = 2k_2\pi$, where $k_1$, $k_2$ and $k_3$ are 
any integers. Prime numbers for $N_c$ are interesting, as there are less resonances satisfying both diophantine conditions simultaneously.


\begin{figure}[pht]
\centering\includegraphics[width=.45\linewidth,height=.32\linewidth]{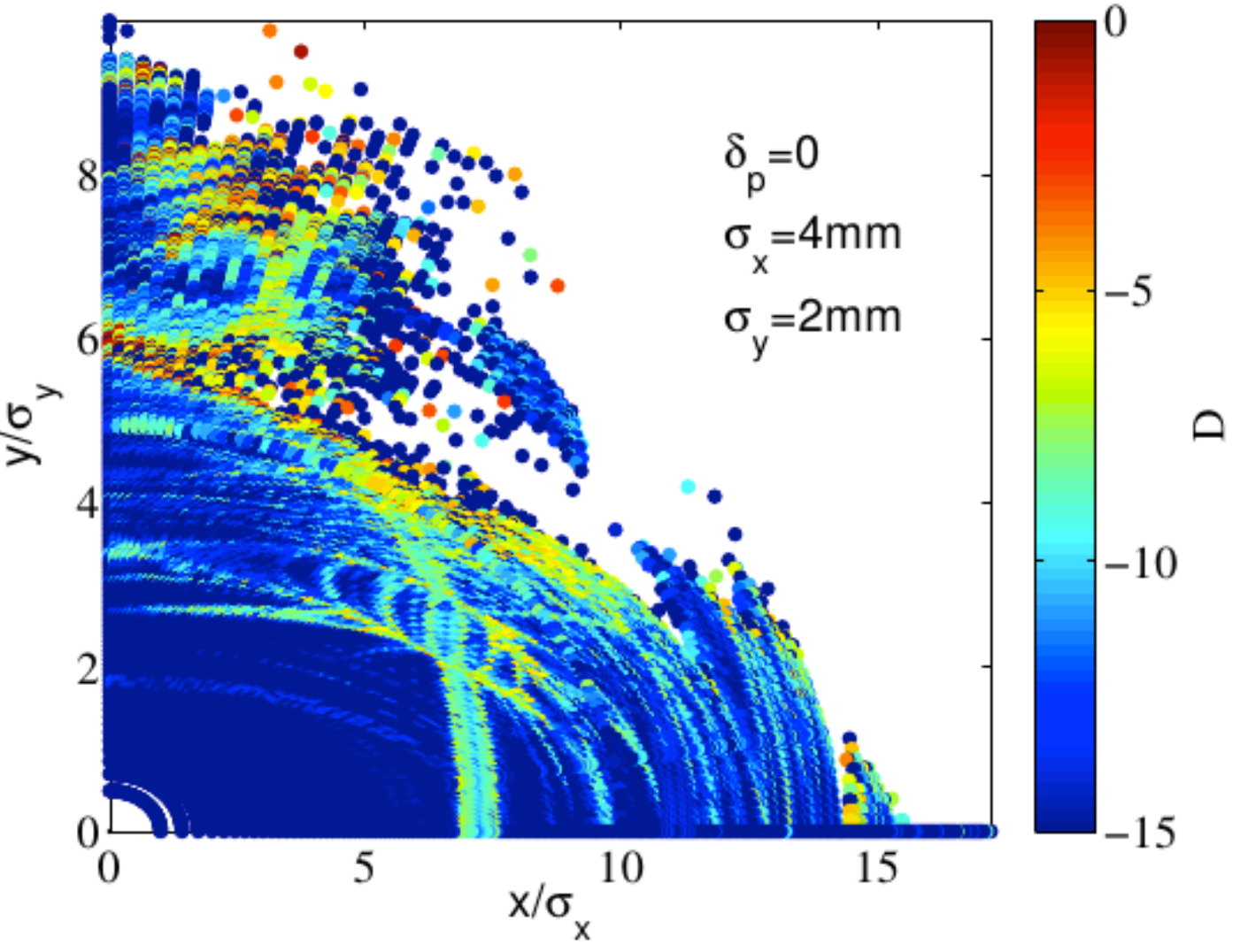}
\centering\includegraphics[width=.45\linewidth,height=.32\linewidth]{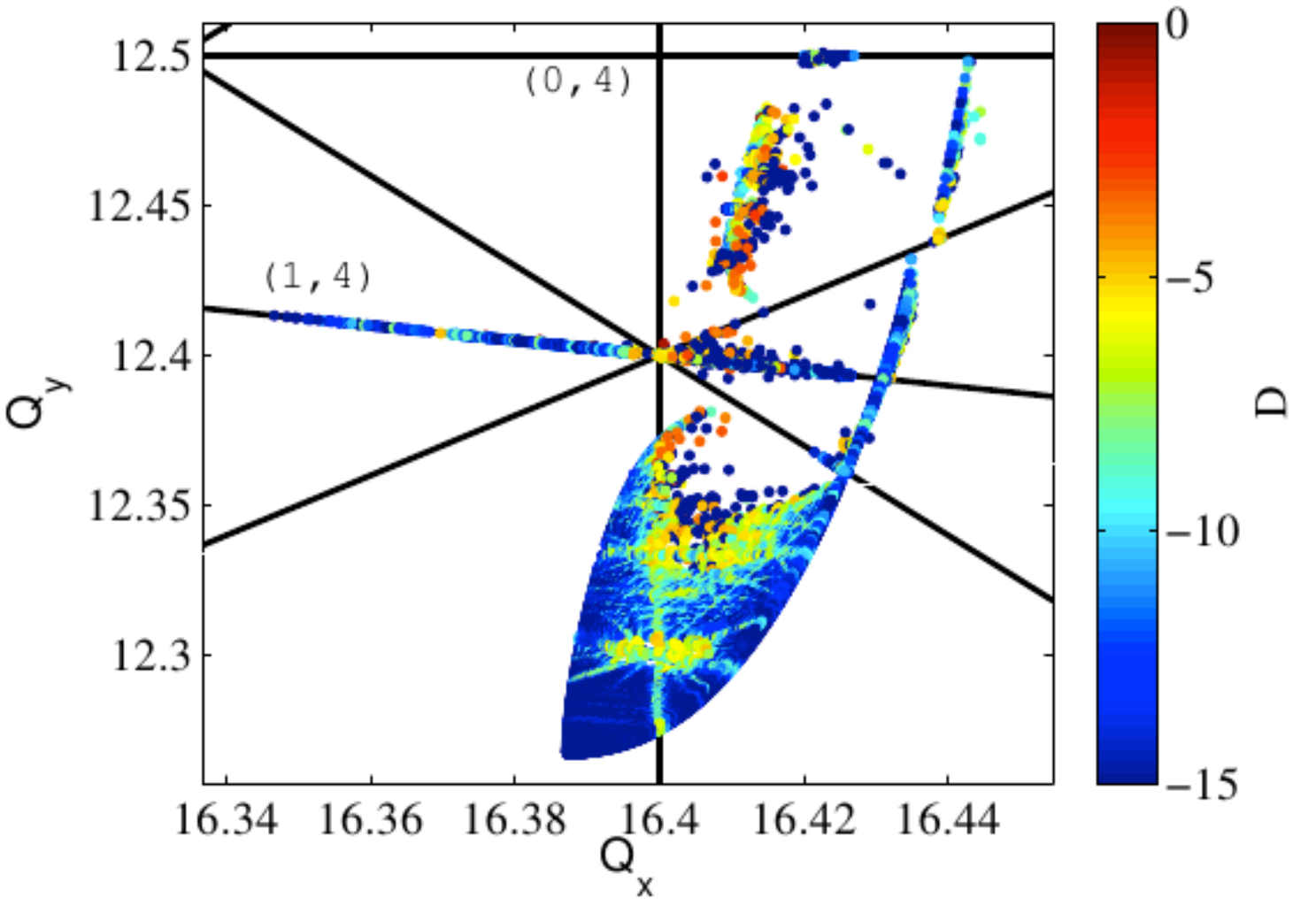} \\
\caption{Frequency maps (right) and diffusion maps (left) for on momentum particles, in the CLIC PDRs, for the working point (16.39,12.26)~\cite{Fanouria}.
}
\label{fig:FM-PDR}
\end{figure}

The nonlinear optimisation of the pre-damping rings (PDR), 
in the injector complex of the future Compact Linear Collider (CLIC),
was based on this ``resonance free lattice concept"~\cite{Fanouria}.
Tracking of particles with different initial conditions for 1024 turns, was performed with MADX-PTC~\cite{ref:madx-ptc},
for a model of the lattice including sextupoles and fringe fields. 
Fig.~\ref{fig:FM-PDR} presents  frequency and diffusion maps for trajectories that survived over 1024 turns, color-coded with the 
diffusion coefficient of Eq.~\eqref{Diffvec}, for on-momentum particles. 
From the frequency maps it is observed that the tune is crossing the (1,4) resonance, which is not eliminated 
by the resonance free lattice and the phase advances chosen ($\mu_x=5/17,~\mu_y=3/17$). 
This seems to be the main limitation of the DA. The shape of the frequency maps, especially at high amplitudes, 
does not have the triangular shape expected by the linear dependence of the tune shift to the action, and 
and foldings appear. This occurs when terms of higher order in the
Hamiltonian become dominant over the quadratic terms, as the amplitude increases. 
This behaviour occurs due to the suppression of the lower order resonances, following the resonance free lattice 
concept.

\section{Experimental non-linear beam dynamics}
\label{expNBD}

\subsection{Experimental frequency analysis}

One of the most impressive aspects of frequency analysis is its ability
to be directly applied in experimental accelerator data. Early studies~\cite{BengtssonPhD} have shown
that the Fourier amplitudes of experimental turn-by-turn (TBT) position data from an accelerator can be associated
to resonance driving terms. In more recent years~\cite{Bartolininormal, TomasPhD, Bartolini2008} and
due to the high precision of the NAFF algorithm~\cite{Laskar88,Laskar90},
frequency analysis of experimental data became quite popular for evaluating and correcting non-linear particle beam
motion. At the same time, it was possible to construct experimental frequency maps~\cite{LaskarPRL00}, 
thereby guiding the experimental calibration of the non-linear accelerator model.  

\begin{figure}[htbp]
\includegraphics*[width=65mm]{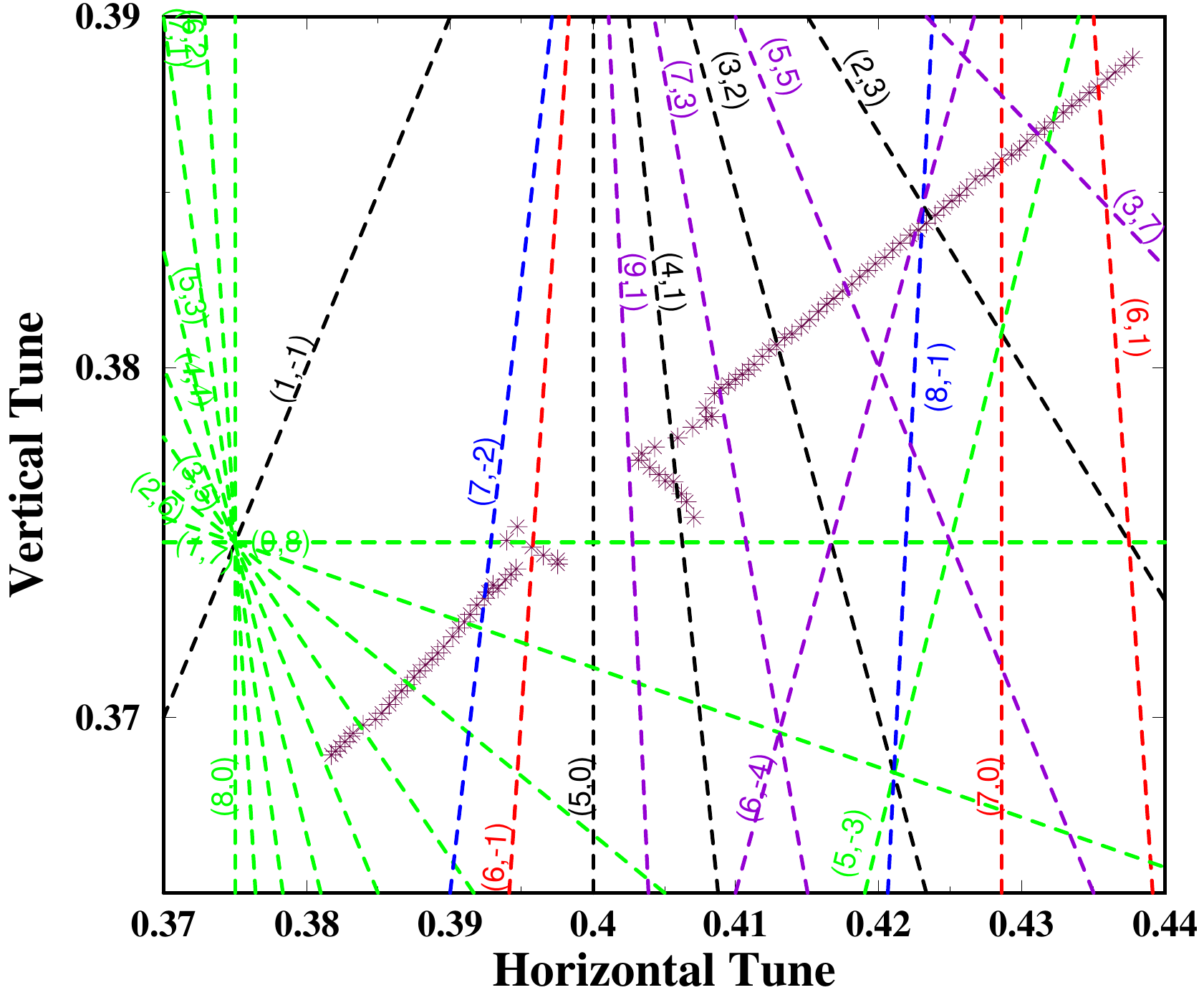}
\includegraphics*[width=65mm]{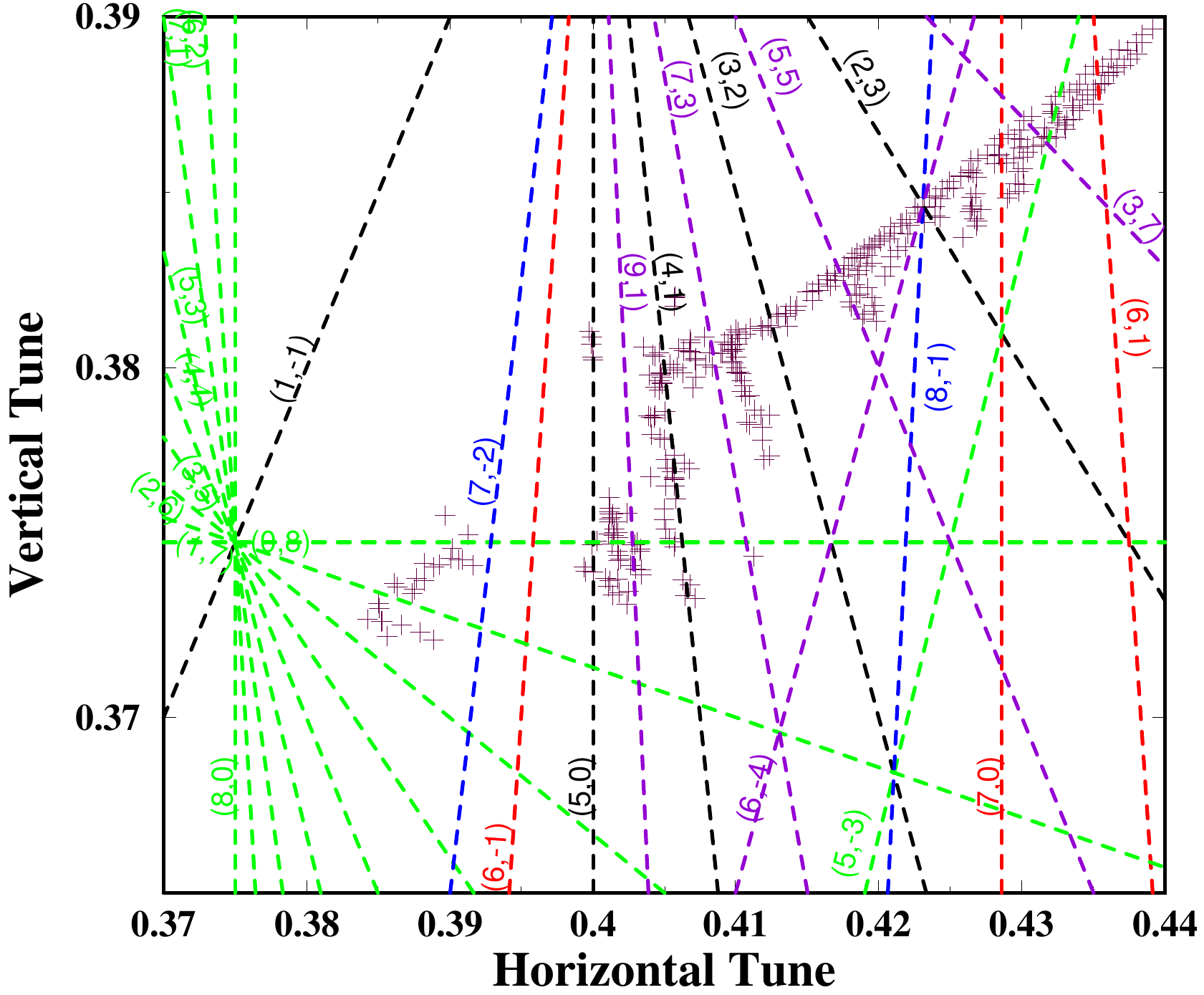}
\caption{Frequency map for the ESRF storage ring for the nominal
sextupole correction (left) and with a detuned sextupole (right), for the working point
(36.44,14.39)~\cite{expfrmap04}.} 
\vspace{-10pt}
\label{frmap1}
\end{figure}

In particular, at the European Synchrotron Radiation Facility (ESRF) storage ring, an experimental
frequency map measurement campaign was initiated~\cite{expfrmap03,expfrmap04},
providing new insight regarding the dynamic aperture limitation of the
operational working point. This can be viewed in the left part of Fig.~\ref{frmap1},
where an experimental frequency map is produced for a large number of horizontal beam excitations and
a constant vertical one and the TBT position data are recorded in beam positions monitors (BPM),
around the ring. The map can be roughly separated in three regions:  For small
amplitudes, the frequency dependence  with the amplitude has
a regular behaviour.  At intermediate  amplitudes, there is
a zone of instability characterised by the accumulation of points on
resonant lines or gaps. Most of the resonances are of the fifth or
tenth order. Finally, for bigger amplitudes, the frequency path is regular, 
crossing a multitude of 8th order resonances that are not excited. The last point, where losses
begin to occur, approaches an area of potential instability, the
crossing point of all third order resonances and the coupling
resonances $(1,-1)$. A frequency map resulting from a numerical
simulation of a model~\cite{expfrmap03}  suggested that the excited zone in the vicinity of the
3rd order resonance is large enough to induce beam loss, as far as the
last measured tunes in this experiment. The same resonance was responsible
for the radical reduction of the DA for momentum deviation of $+2.5\%$~\cite{expfrmap04}.

The right plot of Fig.~\ref{frmap1} represents a frequency map with a detuned sextupole
correction. High order two dimensional resonances appear to be
excited. The frequency space is very distorted, and the dynamic
aperture is limited near the area of the 5th order resonance. This
shows how the frequency map can be used as a
guide to understand the impact of different machine settings in the
beam stability.

\subsection{Loss maps in frequency space}

\begin{figure}[pht]
\centering\includegraphics[width=.45\linewidth,height=.32\linewidth]{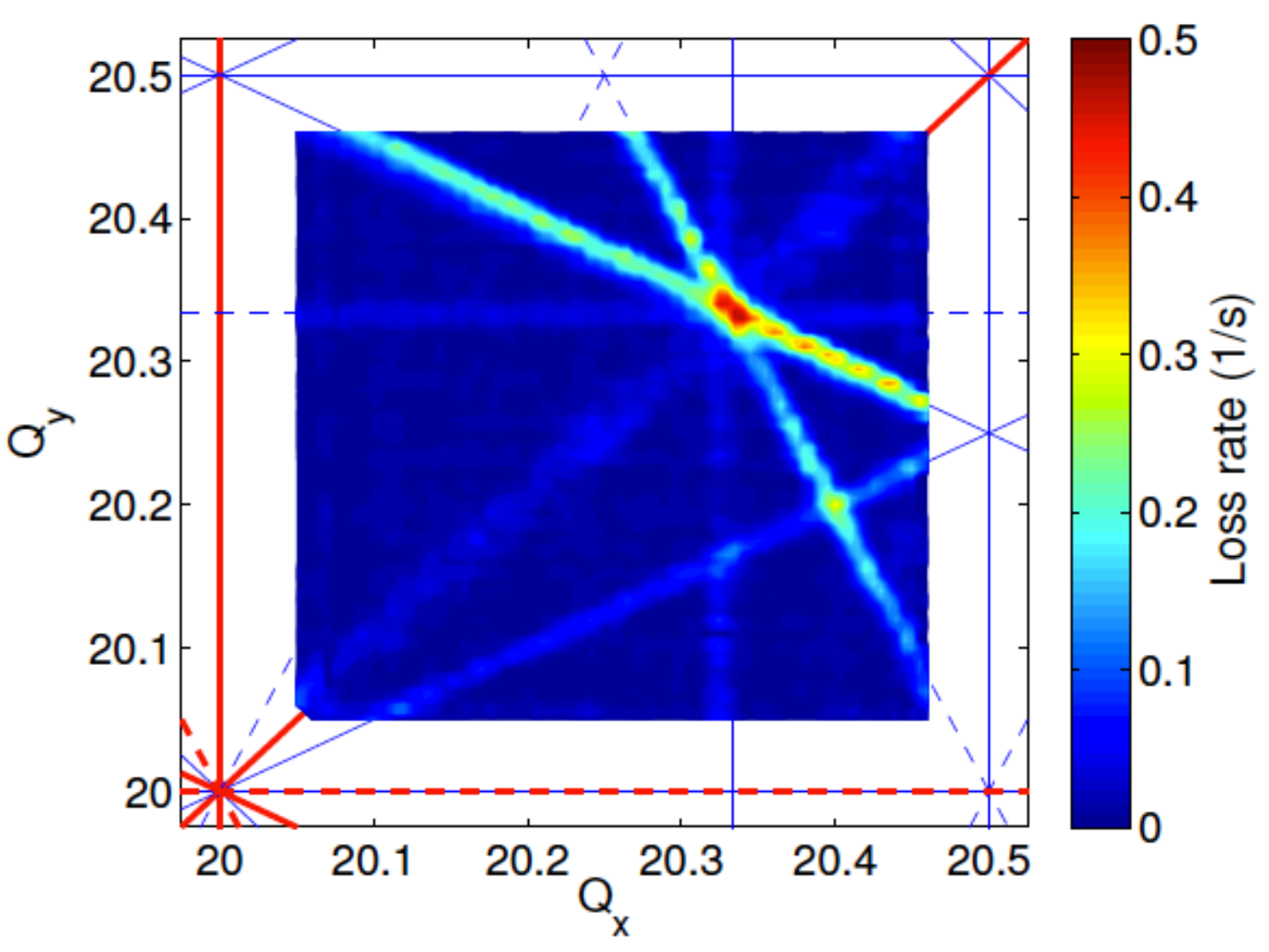}
\centering\includegraphics[width=.45\linewidth,height=.32\linewidth]{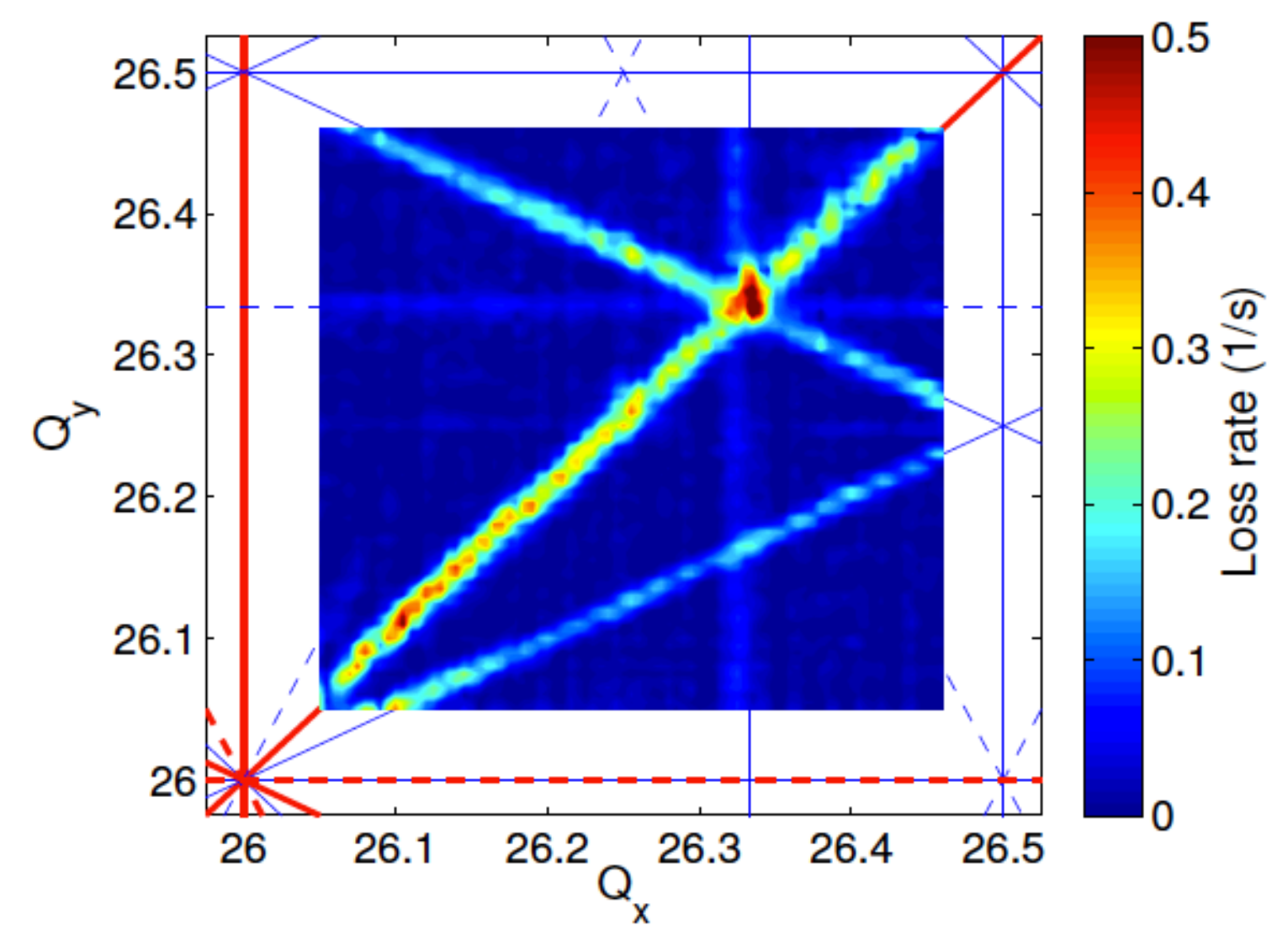} 
\caption{Experimental tune scans in the SPS with the nominal (Q26) optics (left)
and the low $\gamma_t$ Q20 optics (right). The color-code indicates the loss rate
during a dynamic scan of the fractional tunes, as obtained by averaging over 4 scan
directions. 
}
\label{fig:SPSloss}
\end{figure}


Especially in hadron rings, where the frequency variation with amplitude may be quite
small, even up to the physical amplitude limitation, dynamic variation of the  tunes  
for measuring particle losses due to resonance
crossing is a well established technique since decades~\cite{Courant56,Cornacchia76,
CollierLEP, Roncarolo06,franchetti2004, franchetti2005}. 
The measurement principle of the dynamic tune scan used here for comparing resonance
behaviour of  two optics in the Super Proton Synchrotron (SPS)~\cite{Bartosik11} can be described as follows: the beam
intensity is recorded during a slow variation of the betatron tunes. Large particle oscillations 
are induced when the working point is close to a resonance, which results in particle loss. 
The strength of the resonances can be inferred from the slope of the recorded losses as function of time. 
In order to enhance
the observed losses and thus increase the sensitivity to resonances, the beam is injected
deliberately with a large injection error in both planes for provoking transverse emittance
blow-up. The resulting resonance diagrams for  two optics are shown
in Fig.~\ref{fig:SPSloss}, where the colour code indicates the loss rate, 
averaged over the four scan directions. Resonances up to third order can be clearly identified in both
optics. In the case of Q26 optics (left), it appears that the difference coupling resonance $Q_x-Q_y$
was creating higher particle losses compared to the Q20 optics (right).  The normal sextupole
resonances $3Q_x$, $Q_x -2Q_y$ and $Q_x + 2Q_y$ and the skew resonance at $3Q_y$ seem to be
excited in both optics. A surprisingly strong third order skew resonance at $2Q_x + Q_y$
is observed in the Q20 optics.  However,
the area close to the fractional tunes usually used for the LHC beams in the SPS
($0.13,0.18$) is free of strong resonances in both optics. 

\section{Summary}

The study and correction of non-linear effects is crucial for the design and performance 
optimisation of particle accelerators. Non-linear dynamical system methods are fundamental 
in order to achieve this task. The application of the frequency map analysis has become 
a necessary step for understanding and improving the dynamics of accelerator models. Apart
from the global viewing of the dynamics in frequency space and the association of resonant lines to configuration
space, it can provide a frequency quality factor in order to compare different design and correction
approaches. In the case of the LHC, frequency map analysis coupled with normal form construction
and dynamic aperture simulations enabled the identification of dangerous high order  resonances
at injection, guided the choice of correction systems and clearly demonstrated the dominant effect of 
long range beam-beam interactions, at collision.
The method was proven very efficient for choosing the working point of high-power proton rings
like the SNS accumulator and guided the design of a low emittance ring such as the CLIC
pre-damping rings.  The application of the method to experimental data gives great insight 
regarding the dynamic behaviour of  an operating accelerator, such as the ESRF storage
ring or the SPS at CERN. 
Although the method is now well-established for single particle dynamics analysis, it has not yet been applied to study collective particle motion, in a full 6-dimensional, time-varying phase space, with the further inclusion of dissipative effects, due to synchrotron radiation damping. This would be one of the key future challenges which would push the method even beyond its original scope and frame of applicability.

\section*{Acknowledgements}
I would like to express my gratitude to Jacques Laskar for introducing me to frequency
map analysis and the tremendous impact his early mentoring 
had to the rest of my scientific career. I am indebted to  the organisers of the workshop
on ``Methods of Chaos Detection and Predictability: Theory and Applications"
for giving me the opportunity to present this review and in particular to Charis Skokos 
for his long-time friendship and collaboration. Finally, I would like to acknowledge  the contributions
of Fanouria Antoniou, Hannes Bartosik, Frank Schmidt and Frank Zimmermann to a number
of the presented results.

%

\end{document}